\documentclass{aa}  

\usepackage{graphicx}
\usepackage{txfonts}
\usepackage{amsmath}	
\usepackage{amssymb}	
\usepackage{xcolor}
\usepackage{siunitx}
\usepackage{makecell}

\usepackage[]{hyperref}
\hypersetup{colorlinks,citecolor=blue, linkcolor=blue, urlcolor=blue, breaklinks=true}

\begin{document}

   \title{Connecting the growth of galaxies to the large-scale environment in a massive node of the Cosmic Web at z$\sim$3}

    \titlerunning{Connecting the growth of galaxies to the large-scale environment at z$\sim$3}
    \authorrunning{M. Galbiati et al.}

   \author{M. Galbiati\inst{\ref{unimib}}\fnmsep\thanks{\email{m.galbiati29@campus.unimib.it}}
          \and
          S. Cantalupo\inst{\ref{unimib}}
          \and
          C. Steidel\inst{\ref{cali}}
          \and
          A. Pensabene\inst{\ref{unimib}}
          \and
          A. Travascio\inst{\ref{unimib}}
          \and
          W. Wang\inst{\ref{unimib}}
          \and 
          M. Fossati\inst{\ref{unimib},\ref{brera}}
          \and
          M. Fumagalli\inst{\ref{unimib},\ref{oats}}
          \and
          G. Rudie\inst{\ref{carnegie}}
          \and
          A. Fresco\inst{\ref{unimib}}
          \and
          T. Lazeyras\inst{\ref{unimib}}
          \and
          N. Ledos\inst{\ref{unimib}}
          \and
          G. Quadri\inst{\ref{unimib}}
          }

   \institute{Dipartimento di Fisica G. Occhialini, Universit\`a degli Studi di Milano Bicocca, Piazza della Scienza 3, 20126 Milano, Italy \label{unimib}
    \and
    Department of Astronomy, California Institute of Technology, 1200 E. California Boulevard, MC 249-17, Pasadena, CA 91125, USA \label{cali}    
    \and
    INAF – Osservatorio Astronomico di Brera, Via Brera 28, I-21021 Milano, Italy\label{brera}    
    \and
    INAF – Osservatorio Astronomico di Trieste, Via G. B. Tiepolo 11, I-34143 Trieste, Italy\label{oats}    
    \and
    The Carnegie Science Observatories, 813 Santa Barbara Street, Pasadena, California, 91101, USA \label{carnegie}
             }
   \date{\today}

\abstract{A direct link between large-scale environment and galaxy properties is very well established in the local universe. However, very little is known about the role of the environment for galaxy growth before the peak of the cosmic star formation history at $z>3$ due to the rarity of high-redshift, overdense structures. Using a combination of deep, multiwalength observations, including MUSE, JWST, Chandra, HST and ground-based imaging, we detect and study the properties of a population of star-forming galaxies in the field of a hyperluminous quasar at $z\approx3.25$ associated with the giant Ly$\alpha$ nebula MQN01. We find that this region hosts one of the largest overdensity of galaxies found so far at $z>3$, with $\rho/\bar{\rho}=53\pm17$ within $4\times4\rm\,cMpc^2$ and $\lvert\Delta\varv\rvert\leq1000\rm\,km\,s^{-1}$ from the quasar, providing a unique laboratory to study the link between overdense regions and galaxy properties at high redshift. Even in these rare overdense regions, galaxies are forming stars at a rate consistent with the main sequence at $z\approx3$, demonstrating that their SFR is regulated by local properties correlated with their stellar mass rather than by their environment. However, the high-mass-end of the stellar mass function is significantly elevated with respect to that of galaxies in the field at $\log(M_\star/{M_\odot})\gtrsim10.5$, suggesting that massive galaxies in overdense regions build-up their stellar mass earlier or more efficiently than in average regions of the universe. Finally, the overdensity of color-selected Lyman break galaxies observed on larger scales, across $\approx24\times24\rm\,cMpc^2$, is found to be aligned toward the structure traced by the spectroscopically-confirmed galaxies identified with MUSE in the inner $4\times4\rm\,cMpc^2$, suggesting that this highly overdense region could further extend up to a few tens of comoving Mpc.}


   \maketitle
%

\section{Introduction}

According to the current cosmological model, the majority of the inter-galactic gas is distributed in a network of filaments and sheets, called the Cosmic Web, formed through gravitational collapse of the initial density fluctuations \citep{Bond1996}. Structures form in the densest part of the Cosmic Web and grow hierarchically, merging into the massive virialized groups and clusters we observe in today's universe at the intersection of filaments \citep[see e.g.,][]{deLapparent1986, Geller1989, Alpaslan2014, Libeskind2018}. This large-scale environment is thought to play a key role in regulating how the embedded galaxies form and evolve across the cosmic time. Indeed, galaxies found in virialized clusters at $z\lesssim1$ are more massive and form stars at a lower rate compared to the galaxies living in the field \citep[see e.g.,][]{Baldry2006, Bamford2009, Peng2010, Calvi2013, Tomczak2017, Lemaux2019, Tomczak2019, Old2020, Chartab2020, vdBurg2020, McNab2021}. It is thus implied that the galaxies inhabiting extremely overdense regions assemble their mass rapidly and undergo intense episodes of star formation \citep[see e.g.,][]{Thomas2005, Raichoor2011, Gu2018_int}. Therefore, the $z\gtrsim2$ progenitors of present-day clusters, often referred to as protoclusters \citep{Sunyaev1972, Cen2000, Overzier2016, Alberts2022}, are an ideal laboratory within which we can study the fundamental link between galaxies and their large-scale environment.

Overdensities of Ly$\alpha$ and H$\alpha$ emitters, Lyman Break Galaxies (LBGs) or dusty star-forming galaxies are commonly used as a tracer for candidate clusters' progenitors \citep[see e.g.,][]{Steidel1998, Steidel2000, Kurk2000, Kurk2004, Hayashino2004, Chapman2009, Tanaka2011, Lee2014, Casey2015, Oteo2018, Shi2019a, Shi2019b}, although their spatial distribution does not necessarily follow fully virialized structures \citep{Lovell2018, Toshikawa2020}, unlike $z\approx0$ clusters. Indeed, at $z\gtrsim2$, \citet{Chiang2013} predicted protoclusters to extend over $\approx10-20\rm\,cMpc$, while some well studied structures show multiple peaks in their galaxies' overdensity and extend on even much larger scales \citep[see e.g.,][]{Steidel1998, Pentericci2000, Cucciati2018}. In similar $z\gtrsim2$ overdense environments, galaxies are found to be sites of higher star formation activity and their stellar mass function shows an excess at the high-mass-end compared to the field \citep[see e.g.,][]{Kodama2007, Koyama2013a, Koyama2013b, Shimakawa2017, Shimakawa2018, Greenslade2018, Miller2018, Ito2020, Lemaux2022, Taamoli2023, Toshikawa2024} as well as enhanced active galactic nuclei (AGN) activity \citep[e.g.,][]{Lehmer2009, Tozzi2022, Gatica2024}. 
A high rate of mergers \citep{Alonso2012, Mei2023}, efficient gas accretion \citep[e.g.,][]{dAmato2020} and interactions with a {\it proto-}intracluster medium \citep{DiMascolo2023} are expected to play a key role in regulating galaxies growth and activity in these high-redshift overdensities. Though star forming galaxies are observed to be the dominant population in clusters' progenitors \citep{Kuiper2010, Spitler2012, Contini2016}, the detections of quiescent galaxies has been reported in some structures \citep{Kodama2007, Kubo2013, Shi2019b}. These observations suggest that protoclusters could be ideal environments to study the onset of environmental quenching and the processes driving the suppression of the star formation activity observed in $z\approx0$ clusters' galaxies. Theoretical works \citep[e.g.,][]{Muldrew2015, Chiang2017, Lovell2018} are consistent with these observational results, and both support a picture in which galaxies in protoclusters have assembled their masses and rapidly grown at earlier times than those in the field.

Complementary to blind searches, overdense structures of the Inter-Galactic Medium (IGM) itself \citep[mapped via three dimensions Ly$\alpha$ tomography, see e.g.,][]{Lee2018, Newman2020, Newman2022}, as well as the surroundings of highly biased objects, such as High-Redshift Radio Galaxies (HzRGs; \citealp[e.g.,][]{Roettgering1994, Pentericci2000, Hayashi2012, Hatch2014, Rigby2014, Orsi2016, PerezMartinez2023}), Hot Dust-Obscured Galaxies \citep[Hot DOGs;][]{Jones2014, Assef2015, Penney2019, Luo2022, Ginolfi2022, Zewdie2023} and luminous quasars (QSOs; \citealp[e.g.,][]{Kashikawa2007, Venemans2007, Overzier2008, Priddey2008, Kim2009, Utsumi2010, Banados2013, Husband2013, Adams2015, Mignoli2020, Kashino2023}) have been typically explored as signposts of galaxy overdensities. In particular, high-redshift quasars are among the brightest objects in the $z\gtrsim2$ universe and are thus expected to trace overdense structures. Indeed, a hyperluminous quasar at $z\approx2.84$ was found at the center of a large overdensity of LBGs \citep{Steidel2011}, Ly$\alpha$ emitters \citep{Kikuta2019} and sub-millimeter galaxies \citep{Lacaille2019, Wang2024}, tracing a large-scale structure extending across $\approx100\rm\,cMpc$. \citet{GarciaVergara2021} explored the galaxy clustering around $z\approx4$ quasars and found a large overdensity of different galaxy populations, that is CO(4--3) line emitters and Ly$\alpha$ emitting galaxies. On the other hand, \citet{Falder2011, Trainor2012, Adams2015, Hennawi2015} reported the detection of galaxy overdensities in the environment of luminous quasars in only $\approx10\%$ of the cases. 

Despite the increasing number of $z\gtrsim2$ galaxy overdensities and candidate protoclusters discovered in the literature, the heterogeneity of the tracers and the methods adopted to identify the galaxies in their surroundings makes it challenging to draw a unified picture and to consistently trace the assembly and the evolution of the overdense structures down to present-day clusters. Comparing the galaxy environment around radio-loud AGNs from the Clusters Around Radio-Loud AGNs (CARLA; \citealt{Wylezalek2013}) program and radio-quiet quasars with similar masses and redshifts, \citet{Hatch2014} found that the choice of the biased tracer itself has an effect on the large-scale galaxy environment identified in its surroundings. Indeed, on average, radio-loud AGN are observed to be surrounded by larger overdensities and to trace larger structures than radio-quiet quasars. Additionally, differences among overdensities of galaxies selected with a similar criterion around the same biased tracer may arise as a consequence of these structure being in a different evolutionary stage, as shown by, e.g., \citet{Shimakawa2018}.

We present in this work a multiwavelength survey of the large-scale galaxy environment around the $z\approx3.25$ quasar CTS G18.01, which has been pre-selected for being surrounded by extended and filamentary-shaped Ly$\alpha$ emission from the cosmic gas (named MQN01), as discovered by \citet{Borisova2016} and as confirmed later by deeper observations covering a wider area (Cantalupo et al., in prep.). In particular, we targeted the field with a mosaic of observations taken with the Multi Unit Spectroscopic Explorer \citep[MUSE;][]{Bacon2010} at the Very Large Telescope (VLT), the FOcal Reducer and low dispersion Spectrograph 2 \citep[FORS2;][]{Appenzeller1998} and the High Acuity Wide field K-band Imager \citep[HAWK-I;][]{Pirard2004, Casali2006, Kissler-Patig2008, Siebenmorgen2011} at the VLT. The combination of these datasets aims at selecting star forming galaxies at the redshift of the quasar across $24\times24\rm\,cMpc^2$ and provides spectroscopic confirmation within an area of $4\times4\rm\,cMpc^2$ that is covered by MUSE. Additional HST/ACS, JWST/NIRCam and {\it Chandra} observation are used to derive star formation rates and stellar masses by modeling the galaxies' spectral energy distribution (SED) and to explore this peculiar environment in the context of galaxies' formation and evolution. The goal of this work is thus to exploit this rich multiwavelength dataset to explore an overdense region at $z\approx3$ \citep{Pensabene2024} and unveil the roles of both the local and large-scale environment on the stellar mass assembly and star formation activity of the galaxies populating this unique field.

We describe the data in Section \ref{sec:data} and present in Section \ref{sec:selection} the samples of galaxies identified around the target quasar. In Sections \ref{sec:result_zDist}-\ref{sec:radial_overdensity} we investigate the overdensity of star-forming galaxies within $4\times4\rm\,cMpc^2$ around the quasar and explore their properties, compared to galaxies in the field, in Sections \ref{sec:result_MainSeq}-\ref{sec:over_profiles}. We then explore the overdensity of LBGs across $24\times24\rm\,cMpc^2$ in Section \ref{sec:result_Voronoi}. Finally, we discuss this structure in the context of galaxies formation and evolution in Section \ref{sec:discussion}. Throughout, unless otherwise noted, we quote magnitudes in the AB system, distances in physical units, and adopt the Planck 2015 cosmology ($\Omega_{\mathrm{m}} = 0.307$, $H_0 = 67.7$ km s$^{-1}$ Mpc$^{-1}$; \citet{Planck2016}). 


\section{Observations and data reduction}
\label{sec:data}

\subsection{MUSE observations}
\label{sec:data_MUSE} 

The main dataset consists of a mosaic of four VLT/MUSE pointings around the quasar CTS G18.01. The observations are part of a MUSE Guaranteed Time of Observations (GTO) programme ID 0102.A-0448(A), PI S. Cantalupo. They were all conducted in October 2018 in visitor mode with an integration time of, on average\footnote{The final, combined cube consists on a $2\times2$ mosaic of 4 pointings. All of these overlaps in a small region $\approx60\rm\,kpc$ away from the quasar, where the integrated exposure time is of 20h. In all the other regions it ranges from 8h, at the edges, to 12h.}, $\approx10$ hours per pointing using MUSE with AO in Wide-Field mode. The data were taken on clear nights, at airmass $\leq$2.8, angular distance from the moon $\geq30\rm\,deg$, lunar illumination $\leq20\%$ and requiring an image quality better than $\rm FWHM\leq0.8\,arcsec$. The different frames of each pointing were observed with relative inclination of 90 degrees and small dithers ($\lesssim1\rm\,arcsec$) to mitigate the differences in the performance of the 24 IFUs \citep[see e.g.,][]{Borisova2016,Marino2018}. Each of the pointing covers $1\times1\rm\,arcmin^2$, resulting in a final mosaic with a Field-of-View (FoV) of $\approx2\times2\rm\,arcmin^2$ which corresponds to $4\times4\rm\,cMpc^2$ at the redshift of the central quasar, $z\approx3.25$.

The raw data were initially reduced with the ESO Muse pipeline \citep[][version 2.8.1]{Weilbacher2014}. The {\sc scibasic} routine of the ESO Muse pipeline was used to subtract bias, apply flat fielding, perform twilight and illumination corrections, calibrate the wavelengths and the line spread function using arcs and reduce the sky flats. The same calibrations are applied to the field targeting the standard star that is then used to calibrate the flux. The known astrometric offsets \citep{Bacon2015} are corrected by registering the datacubes using the point sources identified in each collapsed individual frame. Once aligned, all the individual frames are combined in a final datacube by using a 3D drizzling interpolation process. The final stacked datacube produced by the ESO pipeline are known \citep[see e.g.,][]{Borisova2016,Bacon2017,Marino2018,Cantalupo2019} to be affected by imperfections in the flat-fielding process and instabilities of the detector. To correct for these effects and improve the quality of the datacubes, we post-process the final products using the {\sc CubExtractor} ({\sc CubEx} hereafter, \citealp[see,][]{DeBeer2023} and Cantalupo et al., in prep.) package, version 1.8, and its subroutines (see figure 1 in \citealt{Lofthouse2020} for an example of the post-processing results compared to the science products reduced by the ESO pipeline). In details, {\sc CubeFix} applies the flat-fielding correction by using the observed sky to self-calibrate the exposures. This process is applied iteratively after masking the continuum-detected sources to avoid contamination. The result is an optimal re-alignment of the relative illumination between the stacks and slices of each IFU, which are not completely corrected by the daily flats. {\sc CubeSharp} performs flux-conserving sky subtraction by including empirical corrections of the line spread function. By means of the {\sc CubeCombine} tool, we then combined the individual exposures into a single datacube. Such cube is then employed to mask continuum sources detected in the collapsed white-light image to optimize a second iteration of {\sc CubeFix} and {\sc CubeSharp}. Finally, {\sc CubeCombine} is used to produce the final and fully-reduced datacube.

\subsection{FORS2 and HAWK-I mosaics}
\label{sec:data_VLT}

\begin{figure} 
\centering
\includegraphics[width=\columnwidth]{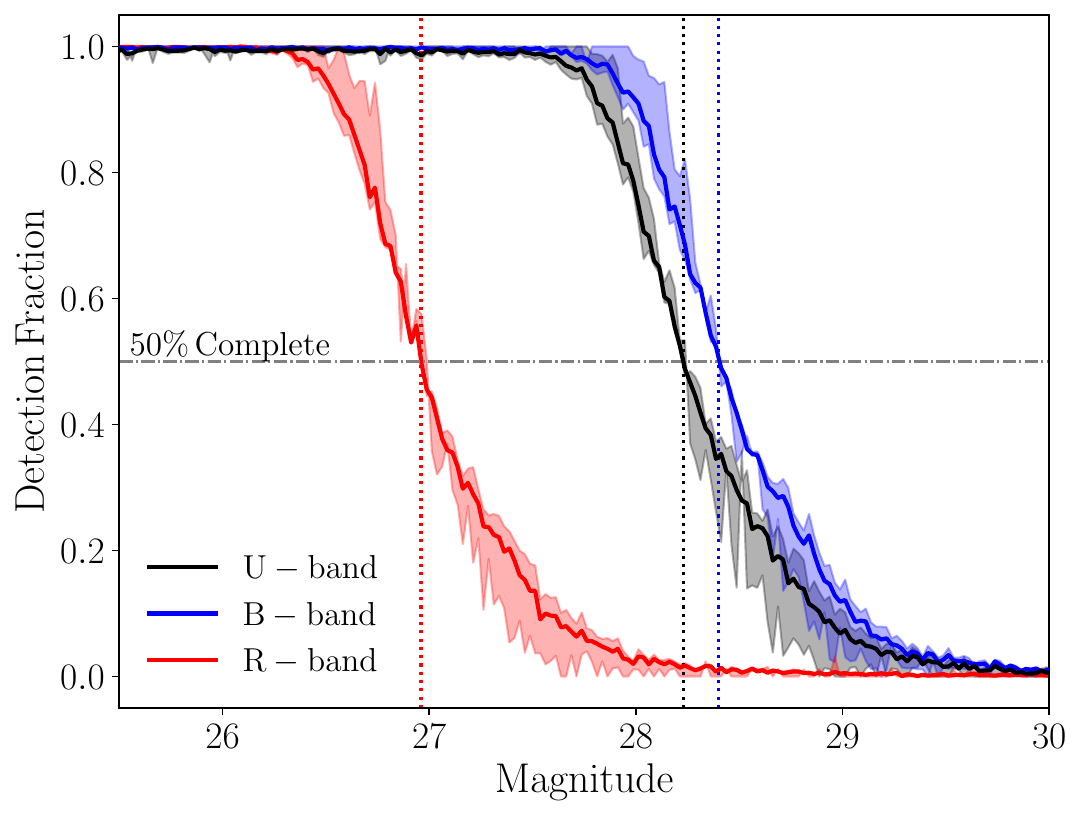}
\caption{Completeness estimated for the VLT/FORS2 images in the U (blue), B (black) and R band (red). The solid line is the result across the entire FoV, while the shaded regions reproduce the variation from the shallower to the deepest regions of each mosaic. The vertical dotted lines mark the magnitudes corresponding to the 50$\%$ completeness limit (black horizontal line).}
\label{fig:data_FORS}
\end{figure}

We followed-up the MUSE observations with a joint VLT/FORS2 and VLT/HAWK-I programme (ID 110.23ZX, PI S.Cantalupo). The FORS2 mosaic is designed to cover an area of $\approx24\times24\rm\,cMpc^2$, which corresponds to a FoV  that is $\approx6\times6$ larger than the one of MUSE and to overlap in the same area covered by MUSE. The observations were executed during ESO Cycle 110-111 in both visitor and service mode, as detailed below and consists of imaging in three different filters: U$_{\rm High}$ (7h on source, $\lambda_{\rm eff}=361\,$nm), B$_{\rm High}$ (5h on source, $\lambda_{\rm eff}=437\,$nm) and R$_{\rm Special}$ (45 minutes on source, $\lambda_{\rm eff}=655\,$nm). The U-band data were taken in visitor mode during two half-nights in October 2022 (Run ID 110.23ZX.001) with the blue-sensitive E2V FORS2 CCD. The B-band and R-band observations were taken in service mode with the MIT CCD. The imaging observations in all the three filters were executed with the default CCD read-out mode for imaging, 2$\times2$ binning and low gain. Our observational constraints include clear nights, airmass $\leq1.5$, $\leq20\%$ of lunar illumination and a seeing better than $\rm 0.8\rm\,arcsec$ FWHM. We observed each mosaic position twice and applied small dithers between the individual frames in each invidual OB to improve contamination removal, background subtraction and to cover the gap between the two chips. To reduce the data, we used the ESO pipeline recipes for FORS2 \citep[][version 5.6.2]{Izzo2010} to apply standard bias, flat and flux calibrations \footnote{The photometric calibration of the B-band and R-band observations is done with the extinction-corrected instrumental zero-point available on the ESO data quality-control database (\url{http://archive.eso.org/qc1/qc1_cgi}). These measurements are processed using the ESO pipeline by combining standard star observations taken for at least 7 stable nights around the nights of the observations. For the U-band images, instead, we run the ESO pipeline on the daily standard frames to measure the zero-point.}. The 2D background was locally modelled and subtracted using the python package {\sc photutils} \citep[][version 1.9.0]{Bradley2023}. Finally, we aligned the astrometry of each individual frame with stars from Gaia DR3 catalog \citep{Gaia2016,Gaia2023} by using the {\sc scamp} package \citep[][version 1.20]{Bertin2006}. The reduced and registered frames are sigma-clipped and combined into the large-scale mosaic using {\sc swarp} \citep[][version 2.21]{Bertin2010}.  

To characterize the depth of the final mosaics, we measured the $3\sigma$ limiting magnitude of each image (see Table \ref{tab:fors}) within apertures of radius $r=4\rm\,pixels\approx1\rm\,arcsec$, corresponding to twice the maximum value of the PSF, and placed in 5000 random positions in blank regions, with all the sources masked to avoid contamination. To estimate the image quality of the science products we used the software {\sc PSFEx} \citep{Bertin2011} which extracts a sample of unsaturated stars to measure the Point Spread Function (PSF) across the entire FoV accounting for spatial variations. We report the mean value across the FoV in Table \ref{tab:fors}, which will also be used to perform accurate photometry later in the analysis: the variation between the minimum and maximum PSF across the FoV is $\lesssim10\%$ for all the images. 

As a final step, we performed injection/recovery experiments to estimate the completeness of the science mosaics. We first created background-only versions of the images by replacing each pixel belonging to known sources, as identified by {\sc SExtractor} (\citealt{Bertin1996}, version 2.24.2) using the same parameters listed in Section \ref{sec:sel_UBR}, with a value randomly extracted from a normal distribution centered on the median value of the background. To create the artificial sources, we convolved point sources with the PSF and randomly injected 1500 mocks for each $\Delta m=0.025\rm\,mag$ bin of magnitude, from $m=24\rm\,mag$ to $m=31\rm\,mag$, in the background-only mosaics. The number of mocks is carefully chosen to avoid overlap and deblending issues. We then run {\sc SExtractor} and measured the fraction of sources we can recover as a function of the magnitude. We compute the results (see Figure \ref{fig:data_FORS} and Table \ref{tab:fors}) for the full mosaic, distinguishing three different regions (the deepest one is designed to overlap with the MUSE FoV at the center of the mosaic) depending on the number of exposures. 

\begin{table}
\caption{Depth and PSF of the VLT/FORS2 imaging observations.}
\centering
\begin{tabular}{ccccc}
\hline
Filter & \begin{tabular}[x]{@{}c@{}} $m_{\rm3\sigma}^{\rm a}$ \\ $(\rm mag)$\end{tabular} & \begin{tabular}[x]{@{}c@{}} $m_{\rm 50\%}^{\rm b}$ \\ $(\rm mag)$\end{tabular} & \begin{tabular}[x]{@{}c@{}} PSF$^{\rm c}$ \\ (arcsec)\end{tabular} \\
\hline
U & 27.90 & 28.25  28.23  28.20 & 0.78  \\
B & 27.45 & 28.40  28.38  28.33 & 0.65  \\
R & 26.08 & 26.99  26.96  26.91 & 0.60  \\
\hline
\end{tabular}
\tablefoot{\tablefoottext{a}{$3\sigma$ magnitude limit measured in apertures of radius $r=4\rm\,pixels\approx1\rm\,arcsec$.} \tablefoottext{b}{50$\%$ completeness limit for the deep, medium and shallow part of the mosaic.} \tablefoottext{c}{Mean PSF (FWHM).}}
\label{tab:fors}
\end{table}

The survey strategy described above for the FORS2 observations was employed to image the same area of $\approx24\times24\rm\,cMpc^2$ around the target quasar with VLT/HAWK-I. We observed the field with a set of three filters: CH4 (1h on source, $\lambda_{\rm eff}=1.575\rm\,\mu m$), H (2h on source, $\lambda_{\rm eff}=1.620\rm\,\mu m$) and K$_{\rm s}$ (2h on source, $\lambda_{\rm eff}=2.146\rm\,\mu m$). The observations were executed in October-December 2022 requiring the same constraints listed for FORS2. The raw data were bias subtracted, flat fielded and flux calibrated by using the ESO Pipeline recipes for HAWK-I imaging (version 2.4.13) implemented in {\sc EsoRex} (\citealt{EsoRex2015}, version 3.13.6). The 2MASS \citep{Skrutskie2006} catalogue is used to measure the flux zero-point. After subtracting the background, we used the {\sc hawki$\_$science$\_$postprocess} recipe in the ESO Pipeline to combine the exposures into the final science mosaic. As a final step, we registered the astrometry of the final mosaic to Gaia by using the {\sc DrizzlePac} package\footnote{\url{https://www.stsci.edu/scientific-community/software/drizzlepac.html}} \citep{Gonzaga2012,Hoffmann2021}.

\subsection{Multiwavelength photometry and ancillary data}
\label{sec:data_ancillary}

\begin{figure*} 
\centering
\includegraphics[width=2\columnwidth]{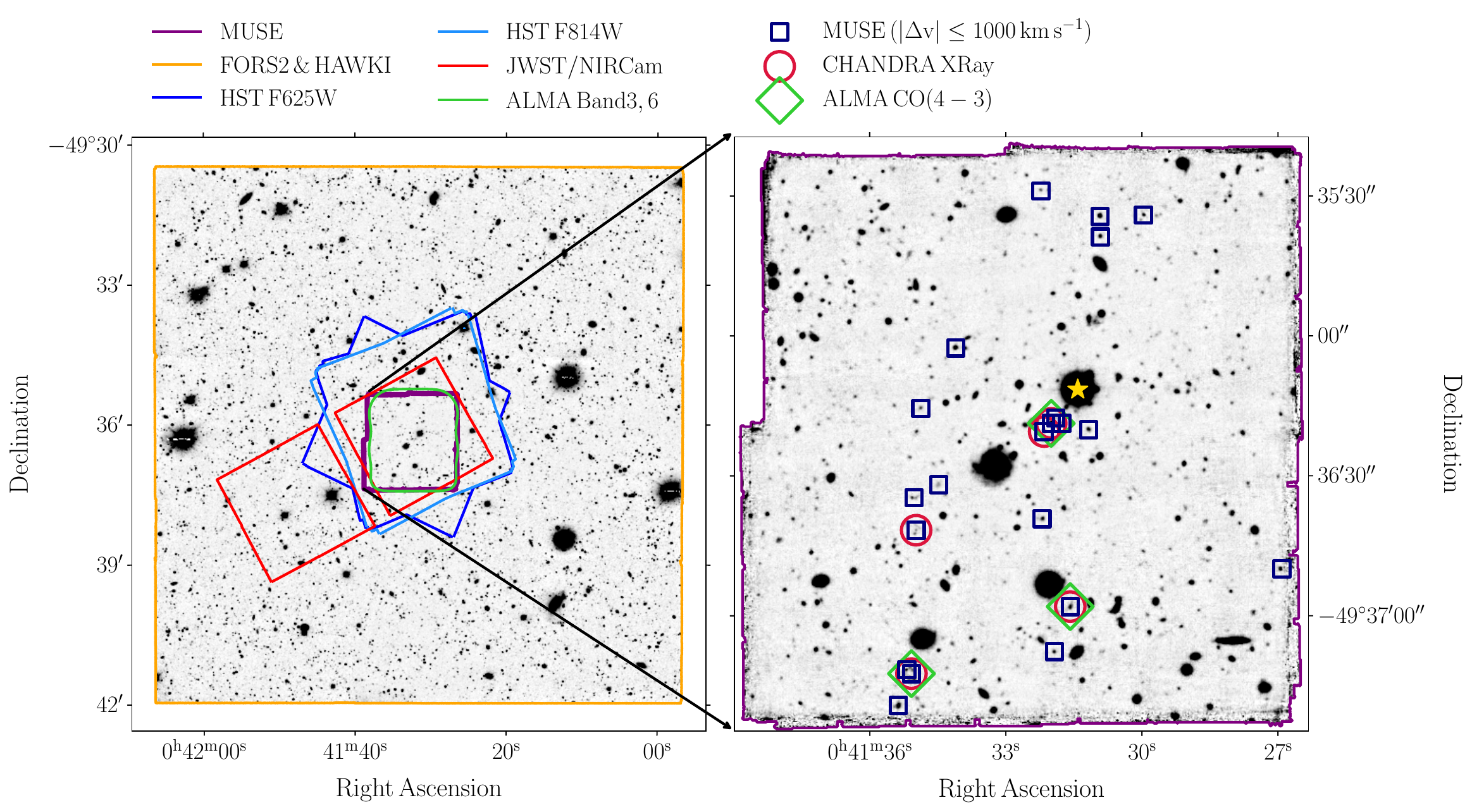}
\caption{Visualisation of the multiwavelength data explored in this work. Left panel: FORS2/R-band image with overlaid the footprint of the MUSE (purple), FORS2 and HAWK-I (orange), HST/ACS F625W (blue), HST/ACS F814W (lightblue), JWST/NIRCam (red) and ALMA/Band3 and Band6 (green) observations. Right panel: the zoom-in shows the MUSE White-Light image in the background with highlighted the MQN01 high-confidence sample of galaxies found in the MUSE FoV within a line-of-sight separation $\lvert\Delta\varv\rvert\leq1000\rm\,km\,s^{-1}$ around the quasar (blue squares), the cross-matched CO(4--3) line-emitters found by ALMA (green diamonds, see \citealt{Pensabene2024}) and X-Ray sources detected by {\it Chandra} (red circles, see Travascio et al., submitted). The target quasar is marked by a gold star.}
\label{fig:Summary_FoV}
\end{figure*}

This field has also been targeted by a multiwavelength campaign spanning from the X-Ray to the millimeter wavelengths (see the left panel in Figure \ref{fig:Summary_FoV}), which enables us to perform a complete census of the different population of sources in the environment of the central quasar, including AGN and dusty star-forming galaxies, that may be elusive in optical observations. Travascio et al., (submitted) collected 600ks of broadband (0.5$\-$7.0 KeV) observations with the Advanced CCD Imaging Spectrometer (ACIS) mounted onboard the {\it Chandra} spacecraft \citep{Weisskopf2000} during Cycle 23 (PI: S.Cantalupo). The final FoV extends above the $24\times24\rm\,cMpc^2$ observed by FORS2 and HAWK-I and provides $\approx1\rm\,arcsec$ resolution imaging and moderate resolution spectroscopy. We also observed the field with the Hubble Space Telescope (HST) for a total of 22 orbits in Cycle 30 (PID: 17065, PI: S. Cantalupo) and used the HST/ACS (Advanced Camera for Survey, \citealt{Ford1998}) instrument to image a $\approx3.4\times3.4\rm\,arcmin^2$ FoV  around the central quasar with the filters ACS F625W (25397 sec) and ACS F814W (30238 sec). We complemented our multiwavelength dataset with the {\it James Webb} Space Telescope (JWST, \citealt{Rigby2023}) pre-imaging observations associated to the GO Cycle 1 programme ID 1835 (\citealt{Cantalupo2021_JWST}, PI S.Cantalupo). With an on-source exposure time of $\approx2\times27\rm\,minutes$, we pointed JWST/NIRCam to image a $2\times5\rm\,arcmin^2$ FoV  with the extra-wide filters F150W2 (in the short wavelength channel, $0.6-2.3\rm\,\mu m$) and F322W2 (in the long wavelength channel, $2.4-5.0\rm\,\mu m$). Details on the data reduction are reported in \citet{Wang2024_bigwheel}. Finally, \citet{Pensabene2024} targeted the field with a Cycle 8 ALMA programme (ID 2021.1.00793.S, PI S. Cantalupo) surveying the $4\times4\rm\,cMpc^2$ MUSE FoV with Band 3 and Band 6 mosaics to detect dusty star forming galaxies via their 1.2 mm continuum emission or CO(4--3) emission lines. The data and the reduction steps are detailed in \citet{Pensabene2024}.


\section{Data analysis}
\label{sec:selection}
 
We describe in this section the methods we adopted to search for galaxies around the target quasar. The rest-frame UV color selections used to identify the Lyman break galaxies in the FORS2 images are detailed in Section \ref{sec:sel_UBR} and Appendix \ref{sec:sel_UGR}. The sample of continuum-selected galaxies identified in the MUSE mosaic, as well as their spectroscopic redshift measurements, is described in Section \ref{sec:sample_MUSE}.

\subsection{Photometric selection of Lyman break galaxies}
\label{sec:sel_UBR}

\begin{figure} 
\centering
\includegraphics[width=\columnwidth]{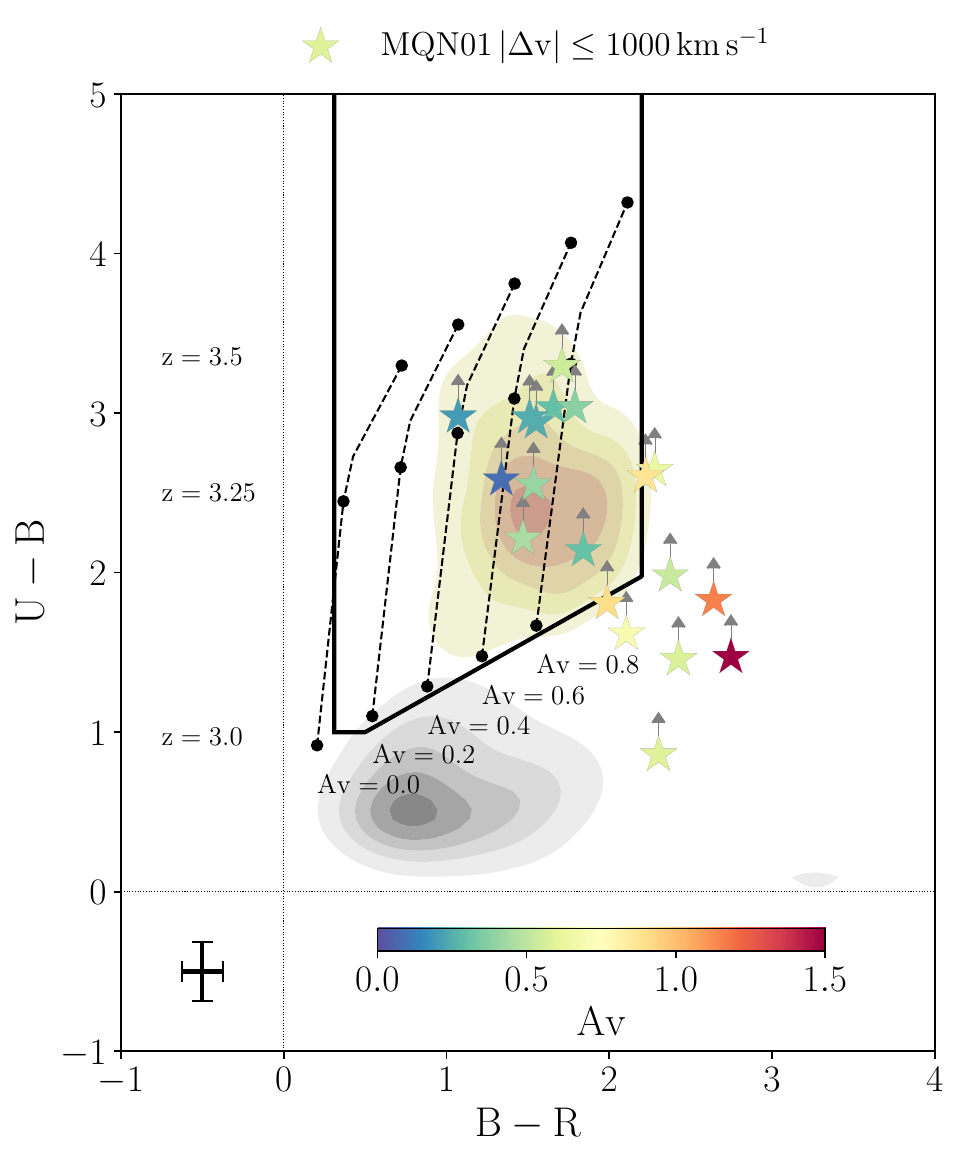}
\caption{$UBR$ color selection of LBGs. The selection region (delimited by solid black lines) is calibrated based on the redshift tracks of typical LBG templates (dashed lines). The orange density plot shows the LBGs candidate at $z\approx3.0-3.5$ that are selected by our criterion. The sources with spectroscopic redshift from MUSE and found within $\lvert\Delta\varv\rvert\leq1000\rm\,km\,s^{-1}$ from the central quasar are shown as stars and are color-coded by their visual extinction, $A_{\rm v}$, estimated by fitting their SEDs (see, Section \ref{sec:SED_fitting}). The gray density plot marks all the R-band detection within FORS2 FoV. Typical errors on the colors are shown in the lower-left corner.}
\label{fig:sample_LBG}
\end{figure}

To identify the Lyman break galaxies at $z\approx3-3.5$ across the FORS2 FoV, we calibrated a selection based on $U-B$ and $B-R$ colors \citep[see e.g.,][]{Bielby2011}. We carefully choose those photometric bands to avoid any possible contamination due to the presence of bright Ly$\alpha$ emission which may be arising from the recombination of the ionizing radiation emitted by the central quasar, rather than from the galaxies' activity. To calibrate the selection region, we used BPASS SPS models (\citealt{BPASS2018}, v2.2-100bin-t8.0-Z001) with a Salpeter IMF \citep{Salpeter1955} up to a stellar mass limit of $M_\star=100M_\odot$, binary evolution and stellar metallicity $Z_\star=0.001$ (corresponding to $Z_\star\approx0.07\,Z_\sun$ with solar abundances from \citealt{Asplund2009}). We assumed a constant star formation history and stellar age of $t=10^8{\rm\,Yr}=100\rm\,Myr$, chosen to be consistent with the median age of the stellar population of the galaxies identified in MUSE (see Section \ref{sec:SED_fitting} and also \citealt{Steidel2018} for a similar approach). We also considered discrete values of reddening: $E(B-V)=0,0.05,0.10,0.15,0.20$, which are typical values for LBGs at $z\approx3$ \citep[see e.g.,][]{Shapley2003}. The models are then post-processed to account for the attenuation due to the absorption from the inter-galactic medium. We thus generated 500 IGM realizations randomly extracted from 1000 sightlines from \citet{Steidel2018} and produced by inserting along each line-of-sight intervening \ion{H}{i} absorption-line-systems, based on their column density distribution function reported in \citet{Rudie2013}, to reproduce the stochastic nature of the IGM absorption. Finally, the color criterion requires:

\begin{enumerate}[(i)]
    \item $U>U_{\rm BKG}^{3\sigma}$
    \item $U-B>1.89$
    \item $B-R<1.74\times(U-B)-1.24$
    \item $0.31<B-R<2.20$
\end{enumerate}

The first condition corresponds to requiring that the source is not detected in U-band above $3\sigma_{\rm BKG}$, where $\sigma_{\rm BKG}$ is the RMS of the background measured within the same aperture used for colors, and is added to avoid contamination from sources at lower redshift and stars (see an application of this same criterion to select U-dropouts galaxies at $z\approx3$ in \citealt{vdBurg2010}). 

We choose the R-band as the detection image and run {\sc SExtractor} with the following parameters: i) {\sc threshold}=1.5 weighted on the mosaic exposure map to assure a uniform threshold across the FoV; ii) {\sc minarea}=5 pixels, corresponding to $\approx1.3\rm\,arcsec^2$; iii) default deblending parameters. The image is also filtered with a $3\times3$ Gaussian kernel to optimize the detection of faint objects. Colors are measured within a fixed circular aperture of $d=5\rm\,pixels\approx1.26\rm\,arcsec$, although we found that the results do not dependent on the radius of the aperture by varying this within the range $0.5-2.0\rm\,arcsec$. The radius is then re-scaled for the variations of the PSF between the different images, as listed in Table \ref{tab:fors}. The results are reported in Figure \ref{fig:sample_LBG}. All the candidates have been visually inspected to remove the objects that lie partially outside the edge of the FoV, the photometry of which cannot be performed on the full-size of the source. As a result, we identified 535 $z\approx3.0-3.5$ LBGs across $24\times24\rm\,cMpc^2$. The quality of the color selections is tested by cross-matching the catalogue of LBG candidates to the spectroscopic sample of galaxies identified in MUSE (Section \ref{sec:sample_MUSE}), as reported in Table \ref{tab:color_plus_MUSE}.

\subsection{MUSE spectroscopic sample of galaxies}
\label{sec:sample_MUSE}

By leveraging the deep MUSE observations, we searched i) for a spectroscopic confirmation of the color-selected Lyman break galaxies; ii) additional continuum-selected sources in close proximity to the central quasar that are not detected in the shallower FORS2/R-band nor color-selected as LBGs. To this end, we run {\sc SExtractor} in the MUSE white-light image, obtained by collapsing the datacube along the wavelength direction, with the parameters: i) {\sc threshold}=2.5 rescaled by the RMS of the background to guarantee a uniform threshold across the FoV; ii) {\sc minarea}=6 pixels, corresponding to $\approx 1.2\rm\,arcsec^2$; iii) minimum deblending parameter {\sc deblend$\_$mincont}=0.0001. The spectra are extracted within an aperture $r=2.5\times R_{\rm Kron}$ to collect the total flux of the sources. We assign a spectroscopic redshift to the continuum-detected sources by using M. Fossati fork\footnote{\url{https://matteofox.github.io/Marz/}} of {\sc Marz} tool \citep{Hinton2016}, which includes additional high-resolution templates suitable for high-redshift galaxies. We attribute confidence scores to the redshift measurements following the method illustrated in \citet{Bielby2019}.

We are able to assign a high-confidence spectroscopic redshift to $\approx38\%$ of the sources. We attribute low confidence to all the sources that exhibit bright Ly$\alpha$ emission at $z\approx z_{\rm QSO}$ on top of a low signal-to-noise continuum and in absence of other spectral features. This choice is motivated by the presence of the bright target quasar, which prevents us from ruling out whether the Ly$\alpha$ emission is powered by a $z\approx z_{\rm QSO}$ galaxy or it arises from recombination emission of the quasar radiation on top of a featureless continuum emitted by a foreground or background object. We thus require at least two of the following stellar and interstellar absorption or nebular emission lines to be clearly recognizable in the rest-frame UV spectra: Ly$\alpha\,1216\,$\AA, \ion{N}{v}$\,\lambda1240\,$\AA, \ion{Si}{ii}$\,\lambda1260\,$\AA, \ion{O}{i}+\ion{Si}{II}$\,\lambda1303\,$\AA, \ion{C}{ii}$\,\lambda1334\,$\AA, \ion{Si}{ii}$\,\lambda1526\,$\AA, \ion{C}{iv}$\,\lambda\lambda1548,1550\,$\AA, \ion{Fe}{ii}$\,\lambda1608\,$\AA, \ion{He}{ii}$\,\lambda1640\,$\AA, \ion{Al}{ii}$\,\lambda\lambda\lambda1670,1854,1862\,$\AA, \ion{C}{iii}$\,\lambda\lambda1906,1908\,$\AA. 

With this method, we identify 21 galaxies with high-confidence spectroscopic redshifts lying within a line-of-sight separation $\lvert\Delta\varv\rvert\leq1000\rm\,km\,s^{-1}$ from the central quasar (right panel in Figure \ref{fig:Summary_FoV}). To prove that our selection of galaxies associated with the QSO is robust against the choice of velocity window,  we searched for galaxies within a larger range $\lvert\Delta\varv\rvert\leq2500\rm\,km\,s^{-1}$ in the MUSE cube and found that the number of sources increases only marginally, by $\lesssim10\%$. In order to complete the characterization the galaxy population around the central quasar, we cross-match the catalogue of MUSE galaxies to the sample of X-Ray emitting AGN observed with {\it Chandra} (Travascio et al., submitted) and to that of dusty star-forming galaxies detected with ALMA by \citet{Pensabene2024}. We identified a total of five AGN detected in X-Ray, three of which are also CO(4--3) line emitters (see Figure \ref{fig:Summary_FoV}). 

As a final step, we cross-match the catalogue of color-selected Lyman break galaxies with the spectroscopic sample assembled in MUSE and report the results in Table \ref{tab:color_plus_MUSE}. The majority of the LBG candidates ($\gtrsim45\%$) is found within $\lvert\Delta\varv\rvert\leq1000\rm\,km\,s^{-1}$ from the quasar. Only $\approx15\%$ has a redshift outside this narrow window, but within the range $2.8<z<3.5$ the color selection is calibrated on. The remaining sources have insufficient spectroscopic information to confidently constrain their redshifts. Compared with the spectroscopic sample, the galaxies that are not color-selected are on average $\times1.2$ more massive (stellar masse $\log(M_\star/M_\odot)\gtrsim10$), and $\times2.4$ more dust extincted (visual extinction $A_{\rm v}\gtrsim0.5$) than those classified as LBGs. 

\begin{table}
\caption{Report of $UBR$ and $UGR$ color selections calibrated to identify $3.0<z<3.5$ Lyman break galaxies around the central quasar.}
\centering
\begin{tabular}{ccc}
\hline
LBGs selection$^{(a)}$ & Area$^{(b)}$ $[\rm cMpc^2]$ & MUSE$^{(c)}$ \\
\hline
UBR  & $24\times24$ & $53\%$ (10/19)  \\
UGR  & $4\times4$   & $79\%$ (15/19)  \\
\hline
\end{tabular}
\tablefoot{\tablefoottext{a}{Color selections described in Section \ref{sec:sel_UBR} and in Appendix \ref{sec:sel_UGR}.}\tablefoottext{b}{Size of the FoV on which the color-selections are performed.} \tablefoottext{c}{Fraction of galaxies with high-confidence spectroscopic redshift within $\lvert\Delta\varv\rvert\leq1000\rm\,km\,s^{-1}$ from the quasar and detected in FORS2/R-band that are independently recovered by the color selections. The 3 sources that are not detected in FORS2/R-band are excluded from this statistics.}}
\label{tab:color_plus_MUSE}
\end{table}

\subsection{Control-field samples}
\label{sec:field_samples}
 
We describe in this section the two independent control samples we built to properly compare the properties of the galaxies identified near the quasar to the expectations for random regions of the universe. To guarantee a consistent comparison, we make sure that these control samples have a similar selection function to the spectroscopic sample of galaxies identified in MUSE. 

We first explored publicly available MUSE datacubes. We selected fields observed with $10$ hours of exposure time and without any bright quasar at $z\approx3.0-3.5$, which are known to be biased tracers of galaxies overdensities \citep[see e.g.,][]{Garcia_Vergara2017}. In the final sample we included the most recent data release \citep{Bacon2023} of the MUSE-Hubble Ultra Deep Field \citep[UDF,][]{Bacon2017}, Q0055$-$269 and Q1317$-$0507 from the MUSEQuBES survey \citep{Muzahid2020}, which were all observed as part of the MUSE-GTO programme, and the archived J162116$+$004823 and J200324$-$325144 \citep{Muzahid2020} from the MAGG survey \citep{Lofthouse2020, Fossati2021, Galbiati2023, Galbiati2024}. To be fully consistent with the selection function of this work, we searched these datacubes for continuum-detected galaxies and required high-confidence spectroscopic redshifts to be measured with the same criterion adopted in Section \ref{sec:sample_MUSE}. As a result, we identify 43 galaxies (hereafter called ``MUSE-FIELD'' sample) with spectroscopic redshift in the range $2.5<z<4.5$ spread across an area over three times larger than the MQN01 MUSE FoV. As a complementary analysis, we also identified a control sample from our own MUSE datacube of the MQN01 including 11 continuum-detected galaxies within the range $2.5\lesssim z\lesssim4.5$, excluding those with velocity separation $\lvert\Delta\varv\rvert\leq1500\rm\,km\,s^{-1}$ from the quasar's redshift to avoid contamination due to the proximity to the central QSO. 

\subsection{Galaxies' physical properties and SED fitting}
\label{sec:SED_fitting}

We aggregated all the photometric measurements from the multiwavelength data described in Section \ref{sec:data} to derive the properties of all the galaxies confidently identified within the redshift range $2.5\lesssim z\lesssim4.5$ in the MQN01 MUSE mosaic, that is, those around the central quasar as well as those in the control-field sample. Specifically, we aim at constraining the stellar mass ($M_\star$), star formation rate ($\rm SFR$) and dust extinction ($A_{\rm v}$) of each source. 

As five of the 21 sources included in the MUSE spectroscopic sample are X-Ray detected AGN, we use the publicly available code {\sc CIGALE} (\citealt{Burgarella2005, Noll2009, Boquien2019}, version 2022.1, which includes X-Ray implementation by \citealt{Yang2020, Yang2022_cigale}) to separate the contribution of the AGN from that of the host and derive a reliable estimate of the galaxies' SFR and stellar mass. We chose high-resolution \citet{BC03} stellar population models, \citet{Chabrier2003} initial mass function and \citet{Calzetti2000} dust law. We fixed both the stellar and the gas-phase metallicity to 0.20 (solar), 0.08 and 0.04 in separate runs. Similarly, we also performed different runs with an exponentially declining, a delayed exponentially declining or an effectively constant ($\tau=10\rm\,Gyr$) star formation history. We notice that such different assumptions on the metallicity and star formation histories lead to results that are consistent within $1\sigma$ with each other, and thus consider the output obtained assuming solar metallicity and exponentially declining star formation history across the paper. The AGN and the X-Ray emissions are modelled following \citet{Fritz2006} and \citet{Yang2022_cigale}, respectively. The full input grid of parameters is adapted from \citet{Barrufet2020} and \citet{Riccio2023}, and listed in Table \ref{tab:CIGALE_AGN}.
To estimate realistic uncertainties on the best-fit stellar mass and SFR, we follow \citet{Pacifici2023} and combine the observed uncertainties inherited from photometry, with those related to the choice of the models (e.g., star formation history and stellar physics). We adopted an uncertainty value of $0.2\rm\,dex$ for the stellar mass \citep[see also,][]{Conroy2013} and of $0.3\rm\,dex$ for the SFR. Compared to the typical errors returned by the fitting code ($\approx 0.09\rm\,dex$ for the stellar masses and $\approx 0.04\rm\,dex$ for the SFR), the dispersion due to systematics is the dominant source of uncertainty. The properties estimated for each galaxy are listed in Table \ref{tab:SED_table}. We also prove that these results are independent on the code used to fit the galaxies' SED in the Appendix \ref{app:SED_table}.


\section{Results}
\label{sec:results}

\subsection{A large overdensity of star-forming galaxies clustered around the quasar}
\label{sec:result_zDist}

\begin{figure} 
\centering
\includegraphics[width=.90\columnwidth]{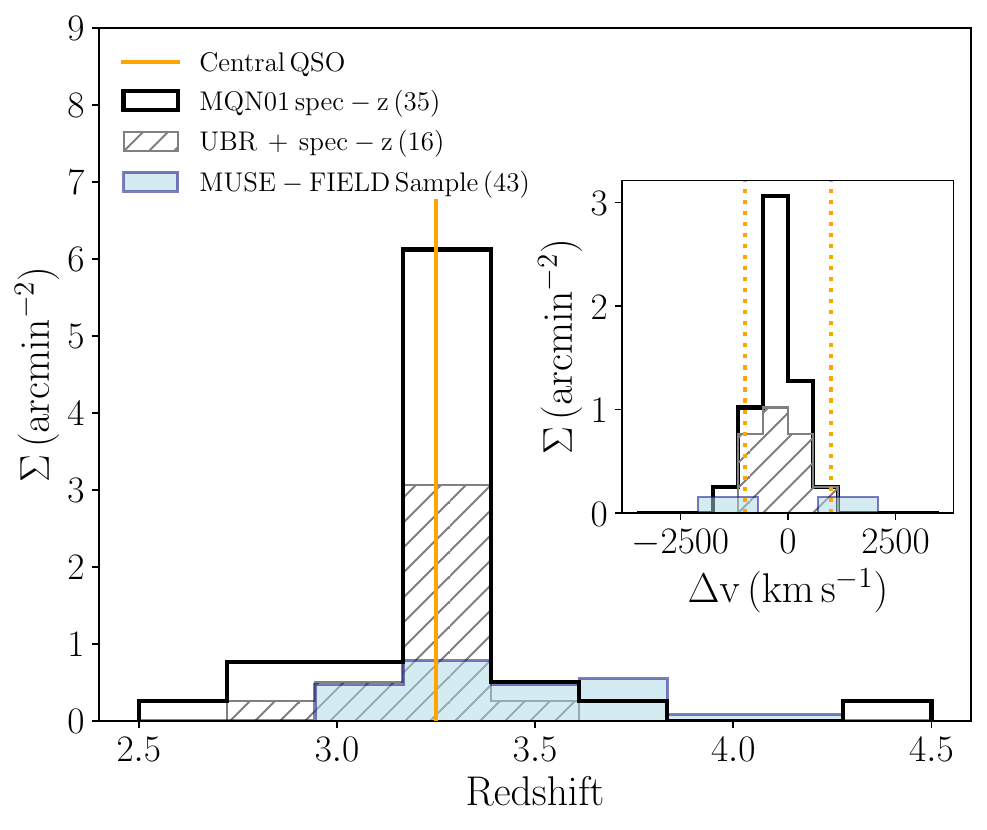}
\caption{Redshift distributions of the MQN01 high-confidence spectroscopic sample (black histogram) and of the $UBR$-selected LBGs with secure spectroscopic redshift from MUSE (hatched gray histogram). These are compared to the ``MUSE-FIELD'' sample and normalized by the surveyed area. The redshift of the central quasar is marked by an orange vertical line. The inset shows the velocity distribution of the MUSE galaxies and of the LBGs relative to the redshift of the quasar. The line-of-sight velocity window corresponding to $\lvert\Delta\varv\rvert\leq1000\rm\,km\,s^{-1}$ is highlighted by dotted vertical lines.}
\label{fig:res_zhist}
\end{figure}

To characterize the large-scale environment around the central QSO, we first investigate the clustering of the galaxies relative to the quasar along the line-of-sight. We reproduce in Figure \ref{fig:res_zhist} the redshift distribution of the MQN01 high-confidence spectroscopic sample of galaxies identified in the MSUE FoV (black) and of the Lyman break galaxies selected via $UBR$ colors (Section \ref{sec:sel_UBR}) for which a secure measurement of the spectroscopic redshift is available from MUSE (hatched gray histogram). The histogram is normalized by the different areas we surveyed to assemble the samples. An excess of galaxies peaked within the redshift interval $3.15\lesssim z\lesssim3.35$ clearly emerges. To roughly estimate the amplitude of the local clustering, we compare these results to expectations for the ``MUSE-FIELD'' sample and report that the surface density of galaxies identified in the redshift bin closest to the quasar is a factor of $\approx8$ larger than the field. Similarly, the surface density of $UBR$-selected galaxies is higher by a factor of $\approx5.5$ compared to the field. Zooming in around the quasar in terms of the line-of-sight velocity, we observe that the majority of the sources are clustered within an even narrower velocity range (see the inset in Figure \ref{fig:res_zhist}), corresponding to $\pm1000\rm\,km\,s^{-1}$. We thus select it as a linking window to identify the galaxies connected to the central quasar throughout the following analysis.

\subsection{Luminosity Function and amplitude of the overdensity}
\label{sec:result_LF}

\begin{figure} 
\centering
\includegraphics[width=\columnwidth]{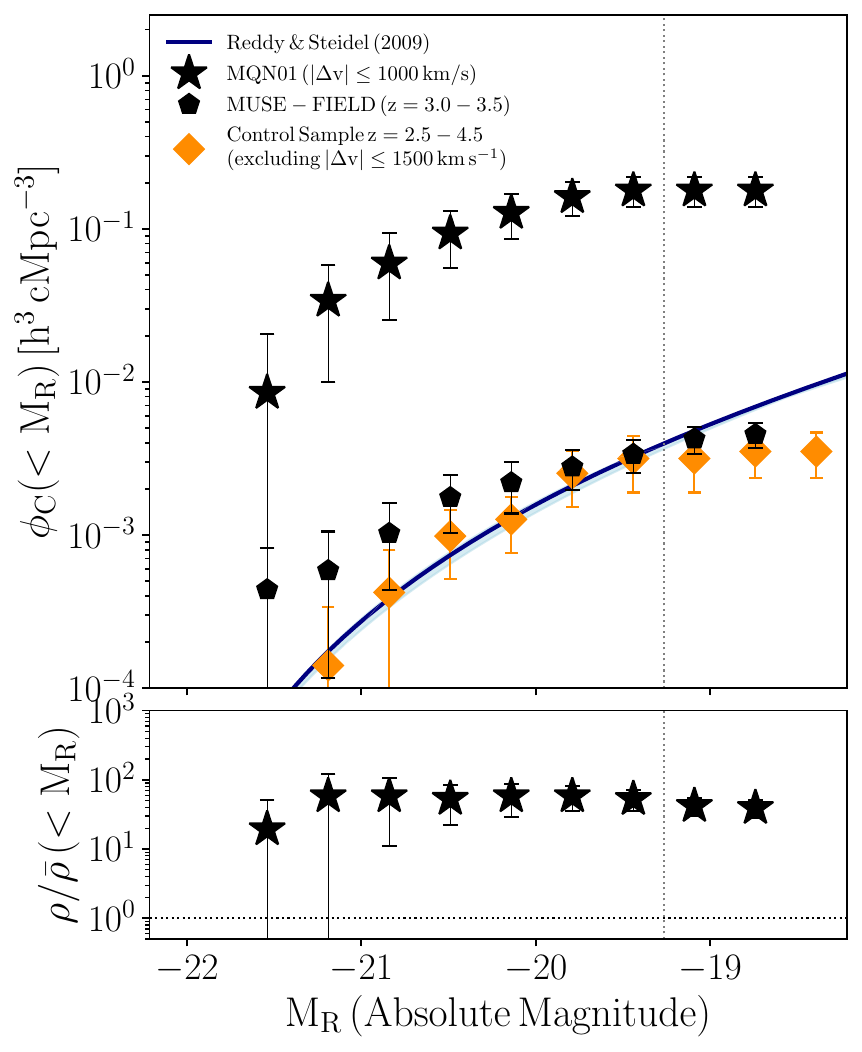}
\caption{Luminosity functions and overdensity of galaxies identified around the central quasar. Upper panel: cumulative R-band luminosity function of the MQN01 galaxies found within $\lvert\Delta\varv\rvert\leq1000\rm\,km\,s^{-1}$ from the redshift of the quasar (black stars), compared to that of the ``MUSE-FIELD'' sample (black pentagons) and of the control sample of sources found in the MQN01 MUSE FoV within $2.5<z<4.5$, masking $\lvert\Delta\varv\rvert\leq1500\rm\,km\,s^{-1}$ around the quasar (orange diamonds). The vertical dotted line show the magnitude at which the luminosity function of the spectroscopic samples flattens relative to the completeness-corrected one from \citet{Reddy2009}. Lower panel: overdensity as a function of the R-band magnitude.}
\label{fig:res_LF_R}
\end{figure}

Once established that an excess of star-forming galaxies is found in proximity to the quasar, we now provide a more statistical argument to investigate the amplitude of the overdensity relative to blank fields. We thus compute the cumulative luminosity function (LF), $\phi_{\rm c}(\rm<M_{R})$, defined as the number density of galaxies as a function of their luminosity. In order to account for the difference in redshift compared to the surveys available in the literature, we convert the flux measured in the R-band into absolute magnitude, $M_{\rm R}$. To obtain the cumulative luminosity function, we first compute the differential number of galaxies at a given magnitude and per unit volume within nine intervals of $\Delta M_{\rm R}=0.175\rm\,mag$ and associate uncertainties from Poisson statistics. The differential measure is then integrated along the magnitude axis. We apply this procedure for the sample of 21 galaxies spectroscopically identified to lie within $\lvert\Delta\varv\rvert\leq1000\rm\,km\,s^{-1}$ around the redshift of the quasar.
The results are shown in Figure \ref{fig:res_LF_R}. We produce the cumulative luminosity function also for the ``MUSE-FIELD'' sample of galaxies in the range $3.0\leq z\leq3.5$ following the same steps described above. For these galaxies, the $M_{\rm R}$ magnitude is estimated by convolving their spectra with the transmission curve of the FORS2/$R_{\rm special}$ filter.
Similarly, we also produce the cumulative luminosity function for all the galaxies identified in the MUSE mosaic within $2.5<z<4.5$, masking the peak of the overdensity in velocity space around the quasar. The luminosity functions of these two independent field samples are consistent with each other, both in shape and normalization. The consistency of the comparison is achieved by having searched for the galaxies in the same MUSE datacube or in those with similar exposure time, $\approx10\rm h$, and applied the same method and constraints to select only the sources with a precise measure of their spectroscopic redshift. On these bases, we assume the selection function to be the same across the samples and do not apply any completeness corrections to the above luminosity functions. In addition, we prove the normalization of the luminosity function of the field sample to be consistent with results derived in \(\mathcal{R}\)-band by \citet{Reddy2009} for a sample of color-selected LBGs with a spectroscopic redshift in the range $2.7\leq z<3.4$. 

To statistically measure the amplitude of the overdensity around the quasar, we take the ratio\footnote{We employed the luminosity functions to measure the expected number density of galaxies for blank fields. For the sample of LBGs, we integrated R-band luminosity function of \citet{Reddy2008} and multiplied it by the completeness curve of the FORS2/R-band (see Figure \ref{fig:data_FORS}) in order to resemble for the selection function of our sample. For the sample of spectroscopically-selected galaxies, instead, such value was directly taken from the luminosity function of the ``MUSE-FIELD'' sample (which is also consistent, within $1\sigma$ uncertainty, to the estimates from the luminosity function of \citet{Reddy2009} and that of the control-field sample we built masking the overdensity around the quasar). As a result, we measured a co-moving density in blank fields to be $\bar{\rho}_{\rm LBGs}=(2.4\pm0.2)\times10^{-3}\rm\,cMpc^{-3}$ and $\bar{\rho}_{\rm spec-z}=(2.5\pm0.8)\times10^{-3}\rm\,cMpc^{-3}$ for the color-selected LBGs and the MUSE spectroscopic sample, respectively.} between the cumulative luminosity function and the one from \citet{Reddy2009}. As shown in Figure \ref{fig:res_LF_R} (lower panel), we measure an overdensity $\rho/\bar{\rho}=53\pm17$ at $M_{\rm R}\leq-19.25\rm\,mag$. We also note that the overdensity is flat as a function of the magnitude threshold up to $M_{\rm R}=-19.5\rm\,mag$ and declines at fainter magnitudes due to the incompleteness of the sample. This suggests that, on average, we only find a shift in the normalization of the luminosity function and do not observe any excess of bright or faint galaxies in the environment around the quasar compared to the field. 
We also perform the same analysis for the sample of color-selected LBGs and show the derived rest-frame UV luminosity functions in Appendix \ref{app:LF_LBGs}.

\subsection{Clustering analysis}
\label{sec:radial_overdensity}

\begin{figure} 
\centering
\includegraphics[width=\columnwidth]{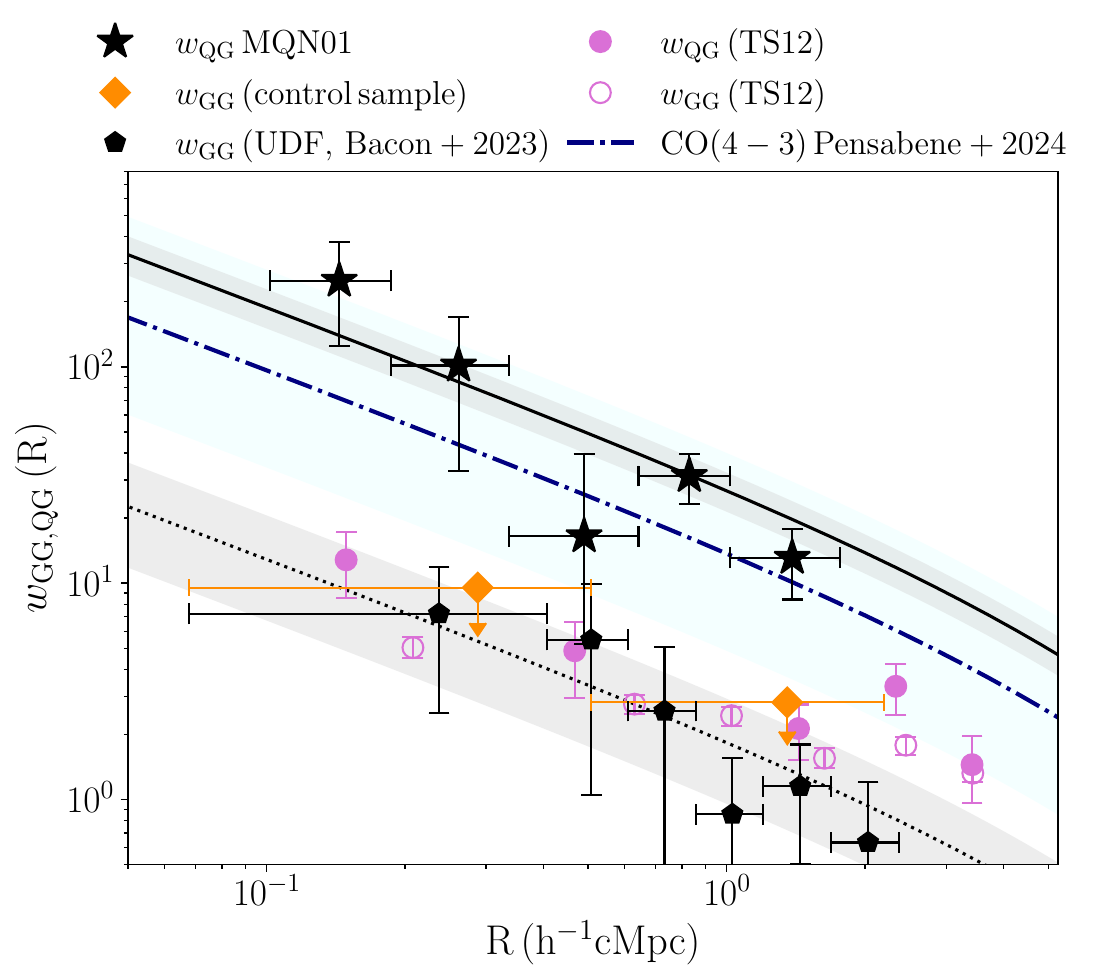}
\caption{Projected correlation functions assuming fixed $\gamma=1.8$ and linking window $\lvert\Delta\varv\rvert\leq1000\rm\,km\,s^{-1}$. The quasar-galaxy cross-correlation function for the MUSE spectroscopic sample (black stars) is similar to that of CO(4--3) line emitters identified by \citet{Pensabene2024} in the same volume (blue line and shaded area), but significantly above that of \citet{Trainor2012} (purple, filled points). The galaxy-galaxy auto-correlation function is shown for the control-field sample of $z=2.5-4.5$ galaxies identified in the UDF \citep[black pentagons,][]{Bacon2023}, together with upper limits for the galaxies in our field within the same redshift range, but masking $\lvert\Delta\varv\rvert\leq1500\rm\,km\,s^{-1}$ around the redshift of the quasar (orange diamonds). Both the estimates are consistent with \citet{Trainor2012} (purple, empty points). The best-fit results are shown as dotted and solid lines for the auto and cross-correlation functions, respectively. All the uncertainties are derived from the 16th and 84th percentiles from bootstrap distributions.}
\label{fig:clustering}
\end{figure}

Comparing the correlation length of the galaxies clustered around different quasars at similar redshifts encodes key information about the structures in which they are hosted. We follow the formalism from \citet{Trainor2012}, TS12 hereafter, and derive the galaxy-galaxy auto-correlation function, $w_{\rm GG}$, and the quasar-galaxies cross-correlation function, $w_{\rm QG}$. In three dimensions, the correlation functions are typically modelled by a power law, $\xi=(r/r_0)^{-\gamma}$, where $\gamma$ is the slope and $r_0$ the correlation length. Once integrated in the redshift direction, the projected correlation functions take a functional form that still depends on the parameters $\gamma$ and $r_0$ (see, eq. 10 in TS12). 
In each radial bin, we compute $w_{\rm GG,QG}(R)=(N_{\rm obs}/N_{\rm exp})-1$, where $N_{\rm obs}$ is the observed number of galaxies identified within a radial bin and $N_{\rm exp}$ is the expected one, derived from the luminosity functions of the field sample from Section \ref{sec:result_LF}. To compute the galaxy-galaxy auto-correlation function without being affecting by the presence of the quasar in our field, we took the $2.5<z<4.5$ galaxies with spectroscopic redshift measured from rest-frame UV absorption lines, that is, similar to our selection in Section \ref{sec:sample_MUSE}, from the UDF catalog \citep{Bacon2023}. In addition, we proved this result to be consistent with upper limits for the auto-correlation of the galaxies identified in our field at $2.5<z<4.5$, masking $\lvert\Delta\varv\rvert\leq1500\rm\,km\,s^{-1}$ around the redshift of the quasar. The results are shown in Figure \ref{fig:clustering}, with the horizontal errorbars marking the width of each radial bin and the vertical ones derived from the 16th and 84th percentile of the distributions obtained by bootstrapping over the sample of galaxies. 
We derive the correlation length by fitting the equation 10 from TS12 to the data. The slope, $\gamma$, is degenerate with the correlation length and therefore fixed to different values to facilitate consistent comparisons with the literature. The best-fit results are shown in Figure \ref{fig:clustering} and listed in Table \ref{tab:clustering}. The uncertainties are obtained from the 16th and 84th percentiles of the bootstrap distributions. The cross-correlation length estimated for spectroscopically-selected star forming galaxies in MQN01 is substantially higher than the galaxy-galaxy correlation lengths obtained using the same selection, and higher than the values found around other quasars in the literature at similar times, e.g. compared to TS12. In Section \ref{sec:disc_overdensity}, we discuss the implications of these results. 

\begin{table}
\caption{Quasar-galaxies cross-correlation and galaxy-galaxy auto-correlation lengths measured for the MUSE spectroscopic sample and for the field galaxies identified in the UDF-MOSAIC \citep{Bacon2023}, respectively.}
\centering
\begin{tabular}{cccc}
\hline
\makecell{Correlation \\ function} & \makecell{$r_0\,[h^{-1}\rm cMpc]$ \\ $\lvert\Delta\varv\rvert\le1000\rm\,km\,s^{-1}$} & \makecell{$r_0\,[h^{-1}\rm cMpc]$ \\ $\lvert\Delta\varv\rvert\le1500\rm\,km\,s^{-1}$} & $\gamma$ \\
\hline \\
$w_{\rm QG}^{\rm MQN01}$ & $23.06^{+3.49}_{-3.43}$ & $22.39^{+3.72}_{-3.36}$ & 1.5 \\[.15cm]
              & $15.71^{+1.79}_{-1.83}$ & $14.08^{+1.76}_{-2.01}$ & 1.8 \\[.25cm]   
\hline \\
$w_{\rm GG}^{\rm UDF}$ & $ 4.25^{+1.56}_{-1.44}$ & $ 2.65^{+1.28}_{-1.10}$ & 1.5 \\[.15cm]
              & $ 3.54^{+1.05}_{-1.08}$ & $ 2.36^{+0.92}_{-0.83}$ & 1.8 \\[.25cm]              
\hline
\end{tabular}
\label{tab:clustering}
\end{table}

\subsection{Galaxies' star formation activity compared to the field}
\label{sec:result_MainSeq}

\begin{figure*} 
\centering
\includegraphics[width=\columnwidth]{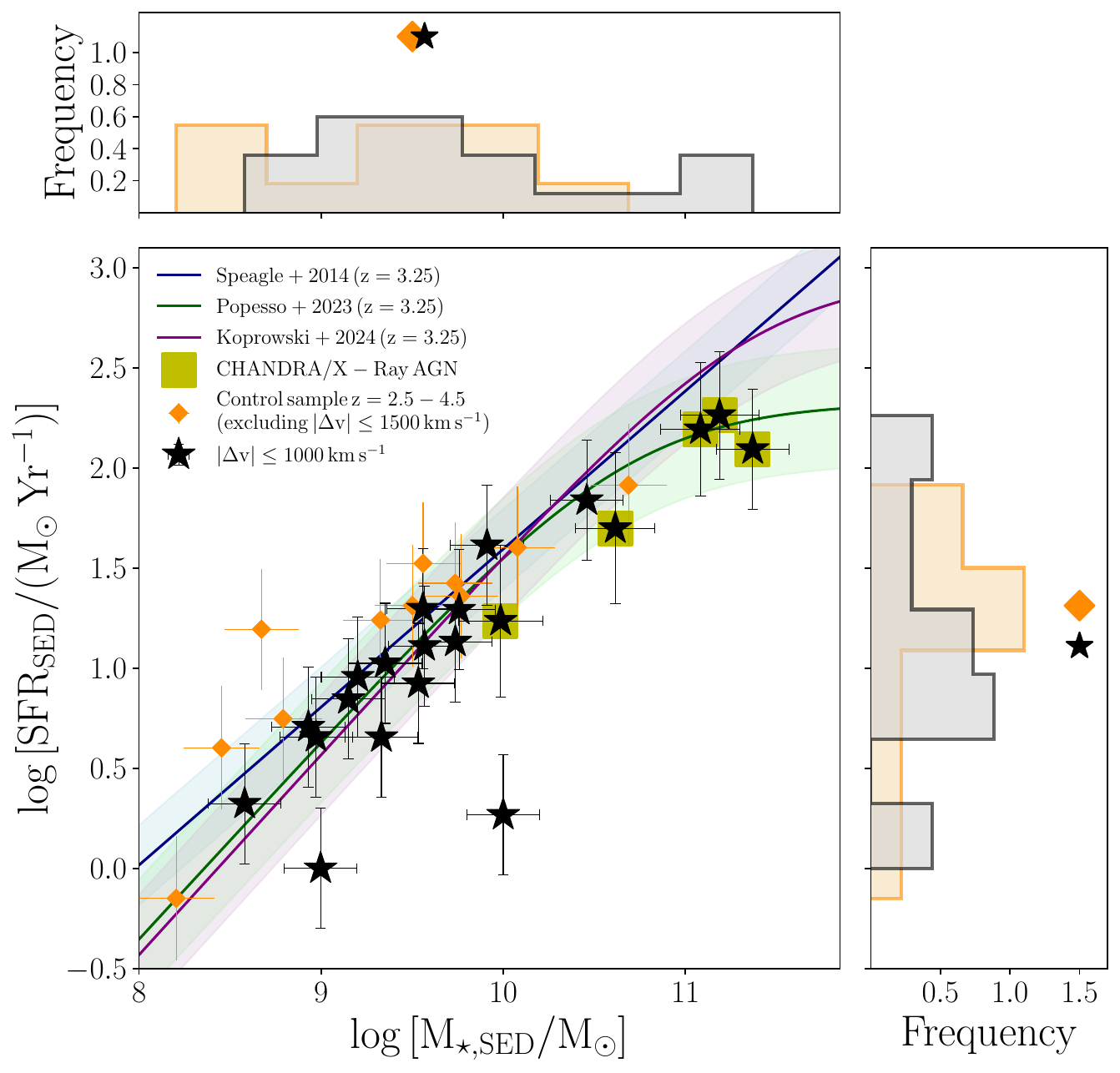}
\includegraphics[width=\columnwidth]{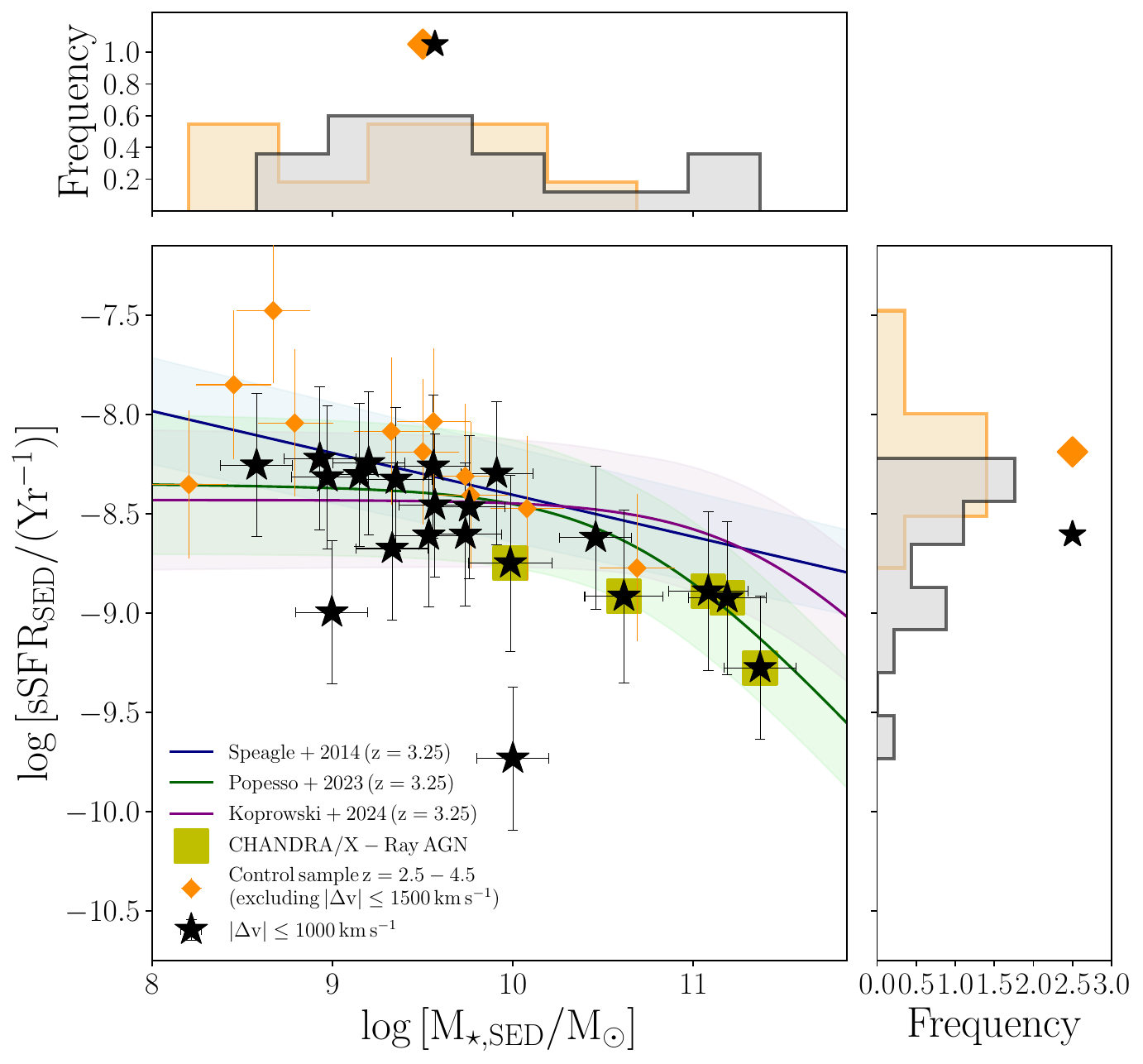}
\caption{Star forming main sequence, in terms of SFR (left panel) and sSFR (right panel), of the MQN01 galaxies clustered around the quasar (black stars), compared to the control-field sample (orange diamonds). The data are compared to the main sequence at z$\approx$3.25 derived from \citet{Speagle2014} (blue), \citet{Popesso2022} (green) and \citet{Koprowski2024} (purple). The X-Ray detected AGN are marked by yellow squares. The histograms on the top and right panels show the distribution of the SFR or sSFR and galaxies' stellar mass, respectively, of the MQN01 galaxies (black) and of the control-field sample, together with their median values on top.}
\label{fig:SED_mainseq}
\end{figure*}

How the large-scale environment affects galaxies growth is a key question within the framework of galaxy evolution in a hierarchical universe, especially in overdense regions throughout cosmic time. A useful approach to gauge galaxies' ability to accrete cold gas and sustain star formation is by looking at their positioning in the stellar mass versus star formation rate (or specific star formation rate, sSFR) plane relative to the "main sequence" \citep[see e.g.,][]{Brinchmann2004, Daddi2007, Elbaz2007, Noeske2007, Salim2007}.

In Figure \ref{fig:SED_mainseq} we directly compare the star formation rate (left panel) and the specific star formation rate (right panel) as a function of galaxies' stellar mass derived for the galaxies clustered around the quasar with the star forming main sequence at $z\approx3.25$. We parametrized the main sequence with the functional forms from \citet{Speagle2014} (blue), \citet{Popesso2022} (green) and \citet{Koprowski2024} (purple) at the quasar redshift, allowing for a $0.3\rm\,dex$ dispersion. 
The method we used to identify the galaxies (Section \ref{sec:sample_MUSE}) selects rest-frame UV bright sources, that is, actively star forming and not too heavily obscured by dust. In other words, our selection tends to exclude passive or extremely red galaxies. As a result, at a given stellar mass, we do not observe any significant offset from the main sequence calibrated on blank fields at $z\approx3$ for the clustered galaxies, nor for the AGNs detected by {\it Chandra} (yellow squares in Figure \ref{fig:SED_mainseq}). The evidence that galaxies form stars at main sequence rates seems to be a recurrent feature in some of the known $z\approx2-3$ protoclusters \citep[see e.g.,][]{Koyama2013a, Koyama2013b, Cooke2014, Shimakawa2017, Shimakawa2018, Shi2019a, Shi2019b, Shi2020, Toshikawa2024}.

Due to the differences in the selection function and in the SFR estimates between the sample of MQN01 galaxies and the ones used to calibrate the parametrizations of the main sequence, it is more informative to directly compare the properties of the galaxies in the overdensity to those included in a field sample with a similar selection function. To this end, we consider the control-field sample of 11 galaxies identified in the MUSE FoV (see, Section \ref{sec:field_samples}). All these galaxies are scattered along the expected main sequence. In addition, we compare the distributions of stellar mass (top panels), SFR and sSFR (right panels) shown in Figure \ref{fig:SED_mainseq} by performing a KS test (significant at $2\sigma$ level for $p$-values$\leq0.05$). While the distributions of sSFR are unlikely drawn from the same parent one, $p-$value$\approx0.01$, there is no statistical difference in those of stellar mass and SFR ($p$-value$\approx0.35$ and $p-$value$\approx0.45$, respectively). Nonetheless, we notice that we find only one galaxy in the field sample located in the $\log(\rm M_\star/M_\odot)\gtrsim10.5$ region of the main sequence, and it is possibly the excess of galaxies in the high-mass tail of the distribution that is driving the difference in the sSFR. The direct comparison between the distributions of the two samples does not take into account the different encompassing co-moving volume ($\approx2\times10^4\rm\,cMpc^3$ for the field sample, which is more than 50 times larger compared to that of the sample of galaxies clustered around the quasar). Normalizing the number of sources by the co-moving volume, we find that the number density of massive ($M_\star\gtrsim10^{10.5}M_\odot$) star forming galaxies is $217^{+200}_{-183}$ times higher around the quasar than in a blank field. Even accounting for the overdensity of $\approx53$ measured around the quasar in Section \ref{sec:result_LF}, the presence of massive galaxies exceeds by a factor of $4^{+2}_{-3}$ the expectations for the field (see, Section \ref{fig:SED_stellarMF} for details). On the other hand, the incidence of $M_\star\lesssim10^{10.5}M_\odot$ galaxies around the quasar is $92^{+80}_{-43}$ times larger than in the field sample, which corresponds to a factor of $1.7^{+1.6}_{-0.8}$ once accounting for the overdensity. This results also corresponds to a factor of $4^{+2}_{-3}$ more galaxies forming stars at a rate $\log[\rm SFR/(M_\odot Yr^{-1})]\geq1.7$ around the quasar compared to blank fields. In terms of sSFR, the massive star-forming galaxies discussed above reside at $\log(\rm sSFR/Yr^{-1})<-8.5$, leading to a large overdensity ($11^{+7}_{-9}$) in this sSFR range as well. This scenario is thus consistent with a picture where a consistent fraction of galaxies in this overdense environment is experiencing a phase of accelerated mass assembly and is rapidly consuming its gas content.

\subsection{Galaxies' stellar mass function}
\label{sec:SMfunction}

\begin{figure*} 
\centering
\includegraphics[width=\columnwidth]{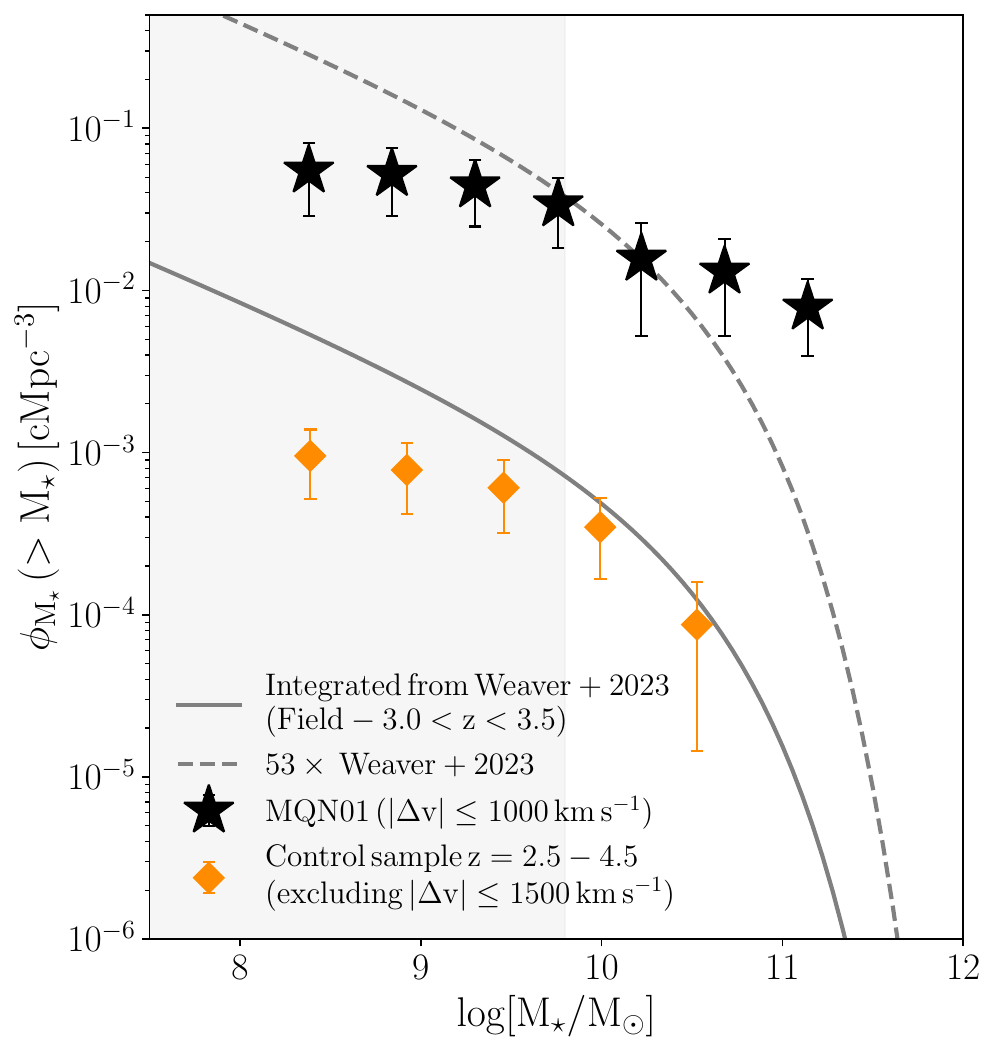}
\includegraphics[width=\columnwidth]{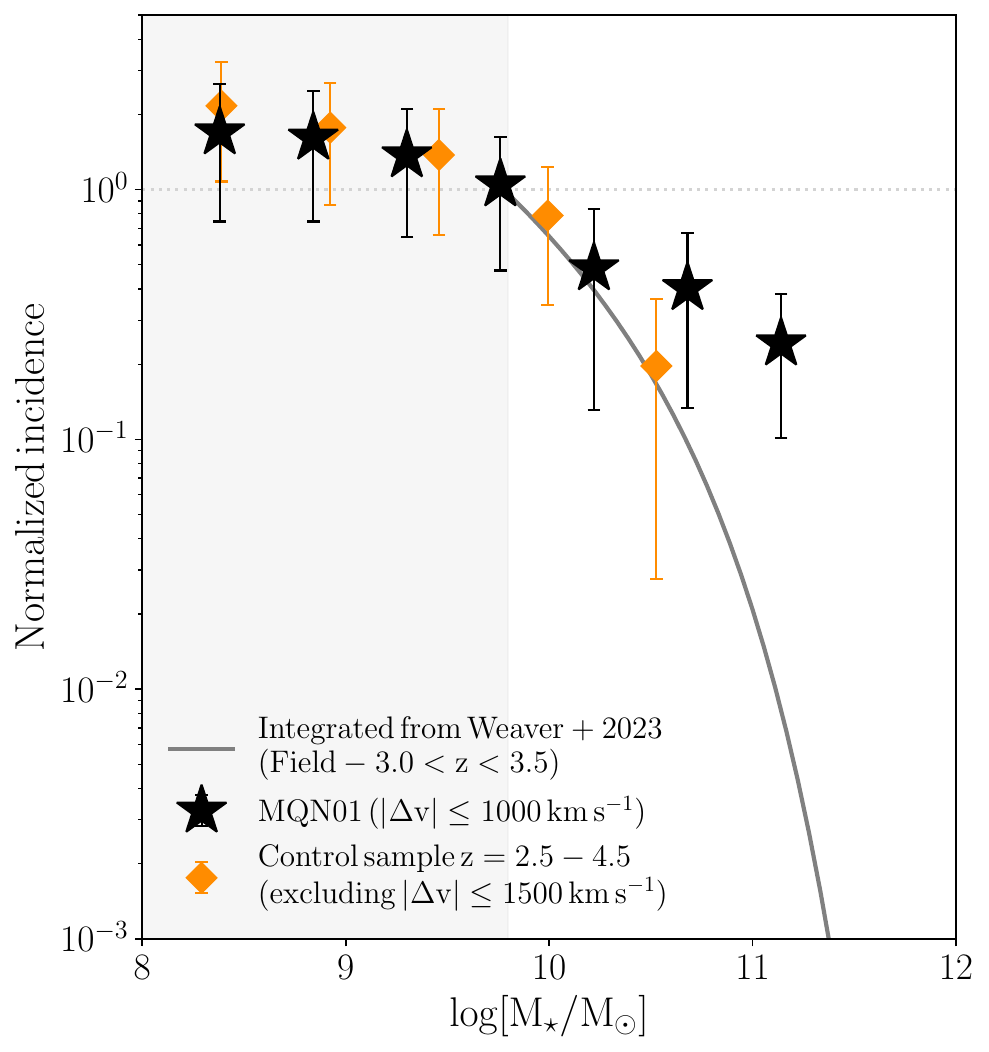}
\caption{Excess of massive $\log(M_\star/M_\odot)\gtrsim10.5$ galaxies around the MQN01 quasar. Left panel: cumulative stellar mass function of the MQN01 spectroscopic sample in proximity to the quasar (black stars) compared to the control-field sample (orange diamonds) and the expectations for blank fields from the COSMOS survey ($3.0<z<3.5$, \citealt{Weaver2023}, gray). Right panel: same as the left panel, but normalized to the value of the cumulative mass function at $M_\star=10^{9.8}M_\odot$, corresponding to the $90\%$ completeness limit of the MQN01 spectroscopic sample.}
\label{fig:SED_stellarMF}
\end{figure*}

Motivated by the results obtained in Section \ref{sec:result_MainSeq}, we now compute the galaxies' number density as a function of their stellar mass by integrating the differential stellar mass function and propagating the Poisson errors arising in each interval of masses. To also take into account the errors on the measure of the stellar mass itself, we bootstrap varying the fiducial values within the uncertainties derived from the galaxies' SED fitting for 1000 times. The standard deviation from the 16th and 84th percentiles of the resulting bootstrap distribution is then squared and summed with errors propagated from the integral above. In the right-panel in Figure \ref{fig:SED_stellarMF} we show the resulting cumulative stellar mass function, while in the left-panel we normalize it at the stellar mass completeness limit\footnote{To estimate an order-of-magnitude completeness for the MUSE spectroscopic samples, we adopt the empirical method from \citet{Pozzetti2010, Davidzon2013, Ilbert2013, Moustakas2013, Laigle2016}, which has already been applied by \citet{Toshikawa2024} on U-dropout galaxies at $z\approx3$. With a K-band magnitude limit of $25.5\rm\,mag$ (see Section \ref{sec:data_ancillary}), we estimate that our sample is 90$\%$ complete at $\log(M_\star/M_\odot)\approx9.8$.} to better capture any difference in shape. The estimate for the galaxies around the quasar (black stars) is compared to the expectations for different field samples. We thus take as a reference the stellar mass function computed by \citet{Weaver2023} in the COSMOS2020 field within the redshift range $3.0<z<3.5$ and rescale it by the MQN01 overdensity $\rho/\bar{\rho}\approx53$. This is also consistent with what we obtain for the galaxies included in the control-field sample and identified within the redshift range $2.5<z<4.5$, masking the peak of the overdensity in velocity space (orange diamonds). While both the stellar mass functions flatten and deviate from \citet{Weaver2023} field expectations at $\log(M_\star/M_\odot)\lesssim9.5$, likely due to the incompleteness of the samples, that of star-forming galaxies clustered with the quasar is also flatter at masses $\log(M_\star/M_\odot)\gtrsim10.5$. In other words, this change in the shape of the stellar mass function provides evidence that the surroundings of the quasar are richer in massive galaxies compared to the field, with $4\pm2$ and $24\pm11$ times more sources in the two bins at $\log(M_\star/M_\odot)\gtrsim10.5$, respectively, once accounted for the overdensity. 

\subsection{Connecting the local overdense environment to the galaxy properties}
\label{sec:over_profiles}

\begin{figure*} 
\centering
\includegraphics[width=\textwidth]{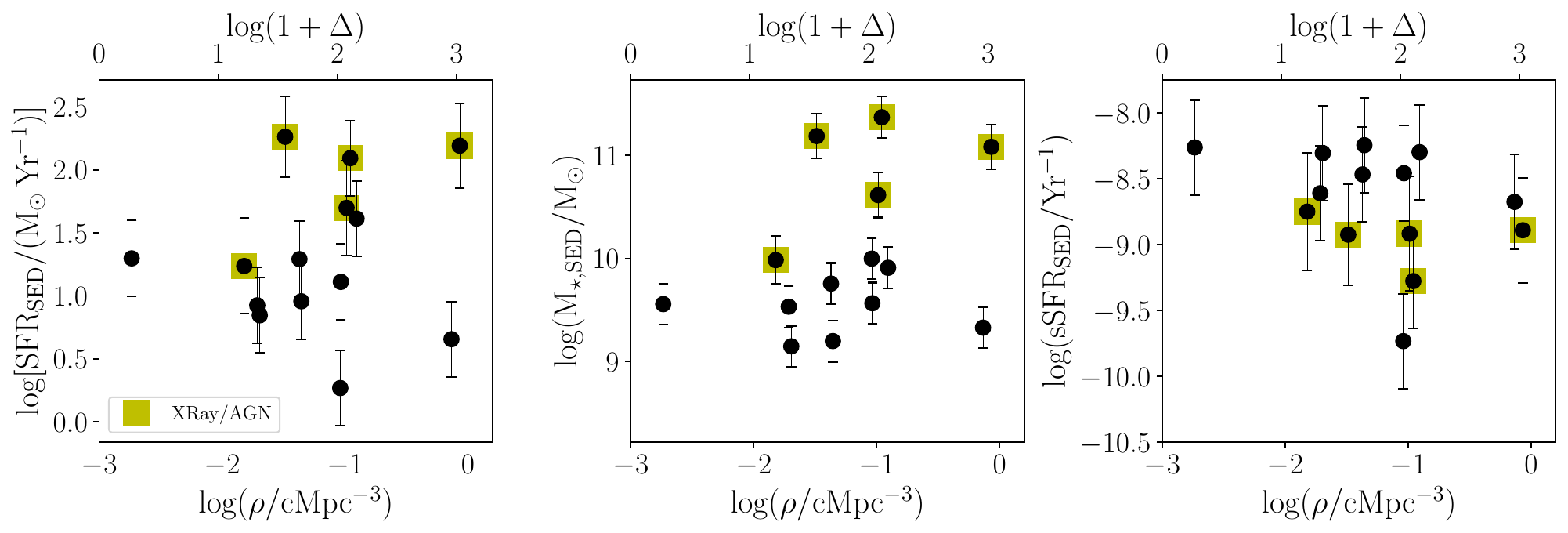}
\caption{MQN01 galaxies’ star formation rate (1st panel), stellar mass (2nd panel) and specific star formation rate (3rd panel) as a function of the co-moving local galaxy density (bottom x-axis) and overdensity (top x-axis) with respect to the field. X-Ray emitting AGN detected in {\it Chandra} observations are highlighted as yellow squares.}
\label{fig:OverdensityProfiles}
\end{figure*}

We now turn to investigate how the properties of the galaxies depends on the local overdensity. We thus employ a scale-independent Voronoi Tessellation (one element per cell) to parametrize the local density of galaxies clustered around the quasar. According to this criterion, the size of each cell is, by construction, inversely proportional to the local galaxy surface density (i.e., small cells corresponds to crowded environments). Open cells at the edge of the FoV (7/21) are excluded from the analysis due to the loss of information about the objects located outside the MUSE mosaic. The overdensity is then defined as $\Delta=\rho/\bar{\rho}-1$ (so that $\log(1+\Delta)=0$ corresponds to the case in which the galaxy density is the same as the field, i.e., $\rho=\bar{\rho}$). We show in Figure \ref{fig:OverdensityProfiles} how the SFR, stellar mass and sSFR correlate with the local galaxy density. To address this analysis on a more statistical basis, we also performed a Spearman-rank correlation test and report in Table \ref{tab:corr_overdensity} the strength of the correlation, $\rho_{\rm spearman}$, and the probability that no monotonic relation exists between the two variables, $p$-value. We require $p$-value$\leq0.05$ for a correlation to be significant at least at 2$\sigma$ confidence level. 
As a result, we observe that the galaxies' stellar mass and specific star formation rate are positively and negatively, respectively, correlated ($|\rho_{\rm spearman}|\gtrsim0.45$ and $p$-value$\approx0.05$) with the local density at $2\sigma$ significance. On the other hand, we do not observe any significant environmental effect on the star formation rate itself ($p-$value$\approx0.29$). 

\begin{table}
\caption{Results of the Spearman-rank test on the correlations between the properties of the MQN01 galaxies and the local overdensity.}
\centering
\begin{tabular}{ccc}
\hline
Variable & $\rho_{\rm Spearman}$ & $p$-value \\
\hline
$\log\rm[SFR/(M_{\odot}Yr^{-1})]$ & +0.27 & 0.29 \\
$\log\rm(M_\star/M_\odot)$ & +0.46 & 0.05 \\
$\log\rm(sSFR/Yr^{-1})$ & -0.50 & 0.04 \\\hline
\end{tabular}
\label{tab:corr_overdensity}
\end{table}

\subsection{Probing the structure on larger scales}
\label{sec:result_Voronoi}

\begin{figure*} 
\centering
\includegraphics[width=.9\columnwidth]{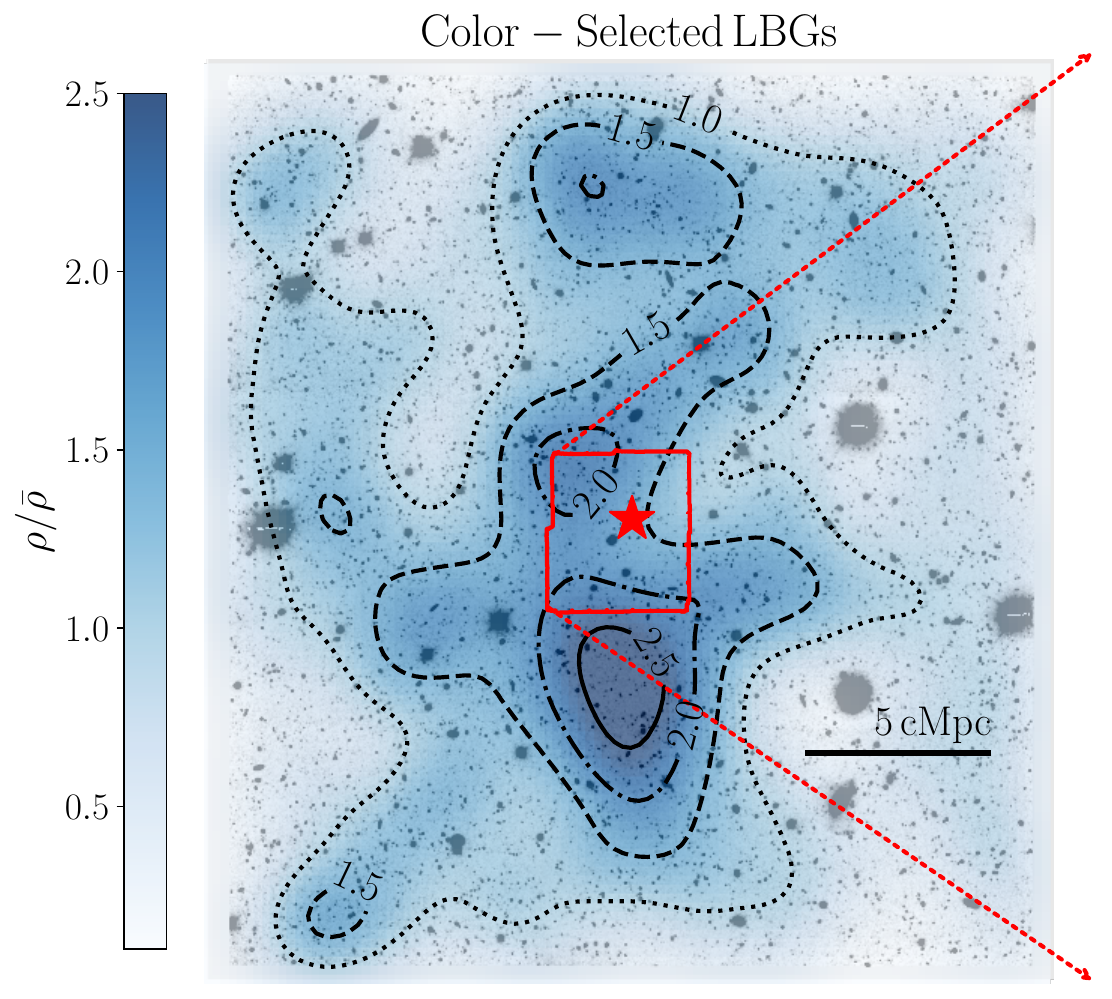}
\includegraphics[width=.9\columnwidth]{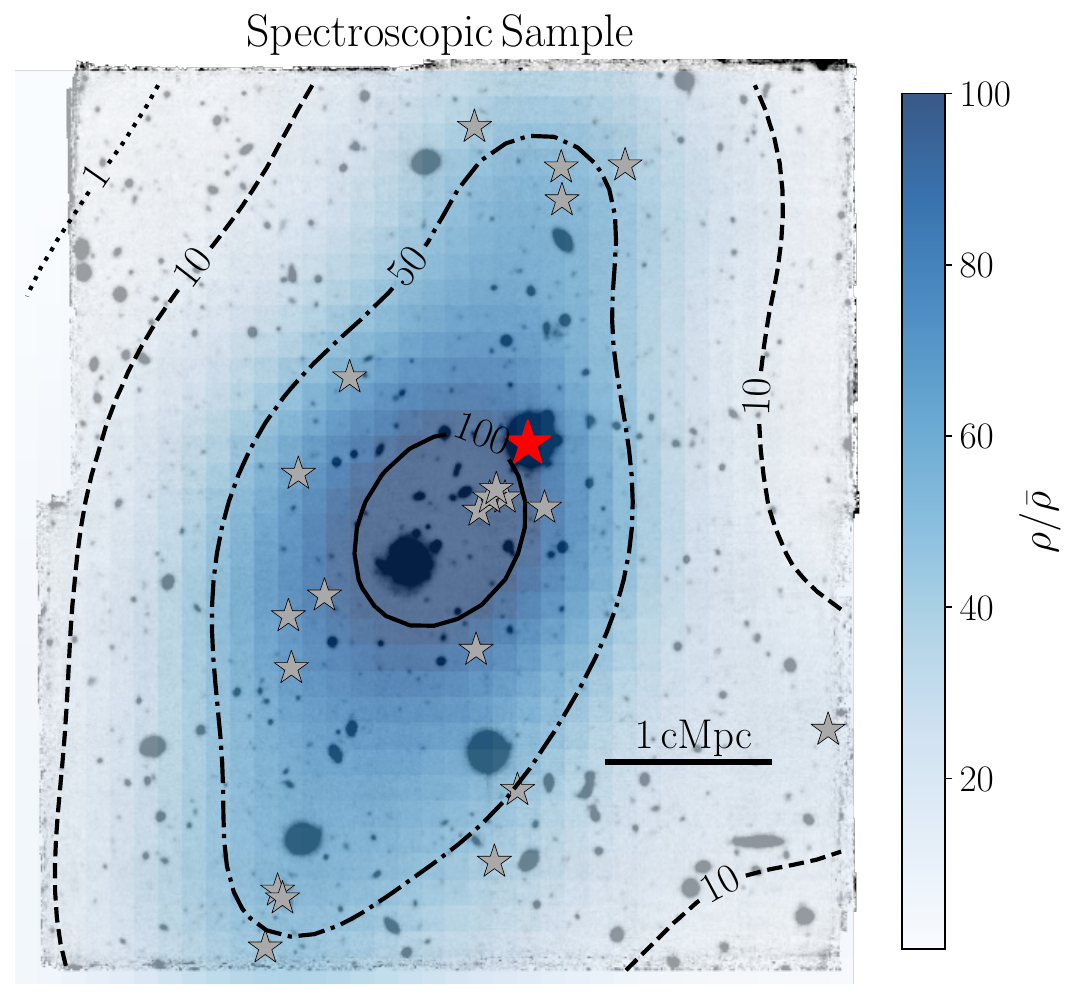}
\caption{2D overdensity maps of the LBGs and of the spectroscopically-selected galaxies found to be clustered with the quasar. Left panel: Overdensity map of the LBGs identified within the $24\times24\rm\,cMpc^2$ FORS2 FoV. The amplitude of the overdensity is shown in shades of blue scaling with the colormap. Contours are shown for $\rho/\bar{\rho}=2.5,~2.0,~1.5,~1.0$ as solid, dash-dotted, dashed and dotted lines respectively. The central quasar is marked as a red star, and the MUSE FoV is also displayed on top of the FORS2 R-band image. Right panel: zoom in the FoV of the MUSE mosaic, with the white-light image overlaid by the overdensity map of the spectroscopic sample of galaxies identified within $\lvert\Delta\varv\rvert\leq1000\rm\,km\,s^{-1}$ (gray stars) around the quasar (red star). Solid, dash-dotted, dashed and dotted contours correspond to $\rho/\bar{\rho}=100,~50,~10,~1$. Note the different spatial scaling and that of the colormap in the two panels. North is up.}
\label{fig:voronoi_LBG}
\end{figure*}

The large-scale structures in which galaxies are embedded are expected to extend well beyond the $\approx4\times4\rm\,cMpc^2$ we are able to trace with the star-forming galaxies identified within the FoV of the MUSE mosaic. We thus model the 2D spatial distribution of the $UBR$-selected LBGs across the $\approx24\times24\rm\,cMpc^2$ FoV  of FORS2 and compare it to that of the spectroscopically selected sources lying within $\lvert\Delta\varv\rvert\leq1000\rm\,km\,s^{-1}$ from the quasar in the $4\times4\rm\,cMpc^2$ FoV of the MUSE mosaic. This is done by creating an evenly spaced grid of $7"\times7"$ and $5"\times5"$ cells covering the entire FORS2 and MUSE FoV, respectively. The number of galaxies identified within each cell is then normalized by the volume of the cell and compared to the co-moving density expected for the field sample (see Section \ref{sec:result_LF}). The grid is then oversampled (by a factor of 3 and 2, respectively) and smoothed with a Gaussian Kernel ($\sigma=6,5$, respectively, corresponding to $\approx0.5"$). Results are shown in Figure \ref{fig:voronoi_LBG}.

The LBGs appear to be overdense ($\rho/\bar{\rho}\gtrsim1$) across the entire $\approx24\times24\,\rm cMpc^2$ FoV, even at large distances from the central quasar. The overdense regions $\rho/\bar{\rho}\gtrsim1.5$ are concentrated closer to the center of the field and show a preferential vertical orientation, which corresponds to the North-South direction on the sky, and is also aligned with both the overdensity of spectroscopically-selected galaxies that we observe in MUSE (right panel) and the extended Ly$\alpha$ emitting gas observed by \citet{Borisova2016} and Cantalupo, et al. (in prep.). This evidence supports a picture in which the large-scale structure traced by the galaxies clustered with the quasar extends well beyond the FoV of the MUSE mosaic, for tens of cMpc. We reserve a more statistical treatment of the LBGs alignment as a function of the local overdensity to Appendix \ref{sec:alignement}. 
On the other hand, when looking at this structure on scales of $24\times24\rm\,cMpc^2$, we observe that the most overdense regions, $\rho/\bar{\rho}\gtrsim2$, are not found within the FoV of the MUSE mosaic, but are mostly concentrated in the southern part of the field. A number of possible scenarios, or a combination of them, should be tested to explain this evidence: i) they are part of a foreground or background structure; ii) the structure surrounding the quasar itself extends in the southern direction; iii) despite being close to the peak of the overdensity in the MUSE FoV, the quasar is not the center of the structure, which is instead in the South; iv) they are tracing two merging or interacting structures. Spectroscopic confirmation of the LBGs would be essential to discriminate between these scenarios.


\section{Discussion}
\label{sec:discussion}

In this section, we put the results presented above into a broader context by comparing them to previous results available in the literature. In particular, in Section \ref{sec:disc_overdensity} we discuss the overdensity and clustering measurements in MQN01 in light of other studies targeting quasars and overdense regions at $z>2$. In Section \ref{sec:disc_galaxy_evolution}, we focus on the galaxies' sSFR and stellar mass function in MQN01 comparing them to other works from the literature, both in the field and in overdense regions, in order to reveal the possible role of the local environment on galaxy formation and evolution. 

\subsection{MQN01 overdensity and clustering compared to other fields}
\label{sec:disc_overdensity}

The number of galaxies spectroscopically confirmed to be within $\lvert\Delta\varv\rvert\leq1000\rm\,km\,s^{-1}$ and in an area of $\approx4\times4\rm\,cMpc^2$ ($\approx2\times2\rm\,arcmin^2$) around the MQN01 quasar results in an overdensity of $\rho/\bar{\rho}=53\pm17$ which corresponds to a surface density of $\approx1.3\rm\,cMpc^{-2}$. At a redshift similar to that of MQN01, several galaxy overdensities and protoclusters have been studied in the literature, such as SSA22 at $z\approx3.08$ \citep{Steidel1998, Steidel2003}, Hyperion at $z\approx2.47$ \citep{Cucciati2018} and Spiderweb at $z\approx2.15$ \citep{Roettgering1994, Pentericci1997}. In particular, \citet{Topping2016} found 40 LBGs in SSA22 with spectroscopic redshift in the range $z=3.05-3.12$ across $17.25\times17.25\rm\,cMpc^2$, which result in a surface density of $\approx0.13\rm\,cMpc^{-2}$. \citet{Huang2022} discovered 53 Ly$\alpha$ emitters in a region extended over $\approx30\times20\rm\,cMpc^2$ in Hyperion, with a surface density of $\approx0.08\rm\,cMpc^{-1}$. In Spiderweb, \citet{PerezMartinez2023} found 39 spectroscopically confirmed H$\alpha$ emitters across $13.7\times19.3\rm\,cMpc^2$, corresponding to a surface density of $\approx0.15\rm\,cMpc^{-1}$. Notably, the surface density measured in the MQN01 field is $\approx10-20$ times higher with respect to these structures. A direct comparison, however, is not straightforward given the different galaxy populations selected in these works, the different levels of completeness reached and the surveyed volumes. A detail comparison would require, for instance, a MUSE (or KCWI) survey in these proto-cluster fields, similar to the one performed around MQN01. 

Besides the large overdensity of star-forming galaxies in the field, a significant excess of galaxies in MQN01 obtained with different tracers has been reported. In particular, \citet{Pensabene2024} found $\approx18$ times more CO(4--3) line emitters and higher 1.2mm source number counts compared to blank fields, with the latter result that is similar to the findings in SSA22. Exploring {\it Chandra} X-Ray data, an overdensity of $\rho/{\bar\rho}\approx1000$ of AGNs has recently been unveiled by Travascio et al., (submitted) and found to exceed that observed in SSA22 and Spiderweb protoclusters. These studies support the peculiarity of the MQN01 field in terms of galaxy overdensity as suggested by our spectroscopic study in the rest-frame UV. 

Stronger evidences about the rarity of the MQN01 structure are obtained by the cross-correlation function presented in Section \ref{sec:radial_overdensity}. In particular, we obtained a quasar-galaxy cross-correlation length of $r_0=15.71^{+1.79}_{-1.83}~h^{-1}{\rm cMpc}$, which is signifcantly higher than the galaxy-galaxy correlation length of $r_0=3.54^{+1.05}_{-1.08}~h^{-1}{\rm cMpc}$ for fixed $\gamma=1.8$. Despite the limited statistics, the cross-correlation lengths obtained using CO(4--3) line emitters around the MQN01 by \citet{Pensabene2024}, $r_0=10.8^{+8.8}_{-4.7}\,h^{-1}\rm cMpc$, are also consistent with these values. These values clearly indicate that galaxies in the MQN01 structure are more clustered around the quasar compared to both the average fields and other quasars at similar redshifts \citep[see, e..g,][]{Trainor2012, GarciaVergara2021}. For consistency and to control for the effects of different selection functions, we compare our results with works that studied the clustering of similar samples of galaxies, within similar volumes and at similar redshifts. TS12 measured an auto-correlation length $r_0=(5.8\pm1.3)\,h^{-1}\rm cMpc$ and a cross-correlation length $r_0=(7.1\pm1.3)\,h^{-1}\rm cMpc$ (once rescaled to the cosmology adopted in this paper, which corresponds to a correction factor of 1.03), assuming $\gamma=1.5$, for a spectroscopic sample of galaxies identified within $\pm1500\rm\,km\,s^{-1}$ from 15 quasars at $z\approx2.7$. The auto-correlation functions of galaxies selected as those found in MQN01, are consistent with that of TS12 (see, Table \ref{tab:clustering}). On the other hand, the cross-correlation lengths of the galaxies identified within $\lvert\Delta\varv\rvert\leq1500\rm\,km\,s^{-1}$ around the quasar is $\approx3.2$ times larger compared to TS12.  
We also measured the quasar-galaxy cross-correlation function in the MAGG survey \citep{Lofthouse2020}, which 28 fields centered on quasars and observed with MUSE (see Appendix \ref{app:clustering_magg} for more details). For consistency with the sample of galaxies used in this work, we considered only those that have been continuum-selected, as described in \citet{Lofthouse2020}, brighter than $m_{\rm R}\le26.5\rm\,mag$, with a secure measure of their spectroscopic redshifts and found within $\lvert\Delta\varv\rvert\le1500\rm\,km\,s^{-1}$ from the redshift of each MAGG quasar (median $z\approx3.7$). We derived a cross-correlation length of $r_0=(3.8\pm0.8)~h^{-1}\rm cMpc$ for fixed $\gamma=1.5$. Interestingly, The one measured in the MQN01 field is $\approx6$ times larger. Altogether, these pieces of evidence suggest that the MQN01 quasar could possibly be associated with a much larger halo than the typical quasar host at these redshifts, which is estimated to be around $\log(M_{\rm h}/M_\odot)\approx12-12.5$ \citep[TS12, see also,][]{Fossati2021, ArrigoniBattaia2022, DeBeer2023}.

\subsection{The role of the local environment on galaxy formation and evolution}
\label{sec:disc_galaxy_evolution}

One of the key results presented in Section \ref{sec:SMfunction} is the fact that the galaxy overdensity in the MQN01 galaxies depends on the galaxy stellar mass. In particular, Figure \ref{fig:SED_stellarMF} shows that there is an elevated fraction of massive galaxies compared to the expectations from average regions of the universe, suggesting a possible link between the environment and galaxy growth. Indeed, despite the limited statistics, we find a hint that the local galaxy overdensity is positively correlated with the galaxies' stellar mass (and negatively correlated with their sSFR) as shown in Figure \ref{fig:OverdensityProfiles}. Our results reinforce previous findings in the literature \citep[see e.g.,][]{Darvish2016, Chartab2020, Lemaux2022, Taamoli2023}, which did not have the large dynamic range in overdensity as provided by our sample. 

For instance, by using a large sample of protocluster candidates, \citet{Toshikawa2024} found that U-dropout galaxies living in overdense regions at $z\approx3$ show a cumulative stellar mass function that is flatter at the massive end compared to the field. Similarly, \citet{Shimakawa2017, Shimakawa2018} found a statistically significant excess of massive galaxies in two $z\approx2$ protoclusters, broadly consistent with other studies of continuum-selected galaxies and H$\alpha$ emitters in overdense regions at similar redshifts \citep[see e.g.,][]{Steidel2005, Koyama2013a, Koyama2013b, Cooke2014}. However, they also observe that such massive galaxies are mostly located in the densest regions of the protocluster, while sources in less dense areas have a stellar mass function distribution that is more similar to that of the field \citep[see also,][at $3.20<z<3.45$]{Forrest2024}. 

The presence of an excess of massive galaxies in overdense regions, such as protoclusters, can be supported by different scenarios: i) the high density environment is able to trigger mergers as well as boost the rate of interactions between galaxies \citep[e.g.,][]{Alonso2012, Cooke2014, dAmato2020}; ii) the access to a large gas reservoir that is efficiently accreted from the cosmic web onto the galaxies \citep[see e.g.,][]{Shimakawa2017}. Indeed, \citet{Umehata2019} found that AGNs and sub-millimeter galaxies in the SSA22 protocluster are located within filaments of gas, which can potentially be accreted by these sources fuelling their star formation and driving their growth. On the other hand, there is also evidence that overdensities play a minor role in shaping the scaling relations of galaxies at $z>2$ \citep[see e.g.,][]{McGee2009, Fossati2017}, which requires further investigation to be conducted in similar regions of the universe. Altogether, these possible scenarios suggest that these galaxies grew rapidly, either assembling their masses earlier or more efficiently than those in the field. 

\subsection{Caveats and future prospects}
\label{sec:disc_limitations}

The most important step for a more complete view of the galaxy population in MQN01 is the inclusion of the passive galaxies. One possibility to select them would require, for instance, properly sampling the Balmer break, expected to be in the H-band at the MQN01 redshift. Photometric data obtained with VLT/HAWK-I are currently available and will be used in a future study to complement the dataset presented here. In addition, JWST spectra covering the Balmer lines at rest-frame wavelengths from 400 to 700 nm are available for some galaxies in the field \citep[][]{Cantalupo2021_JWST, Wang2024_bigwheel}. However, for a more complete view of the passive galaxy population grism JWST spectroscopy would be necessary. 
Only a small area of the MQN01 structure, as probed through our LBG survey, currently has MUSE coverage and thus spectroscopic confirmation. An ongoing MUSE survey (Program ID 112.25N3, PI S. Cantalupo) is covering $\approx14\rm\,cMpc$ in the North-South direction and will be used to extend the study of the clustering and the galaxies' properties up to the outskirt of the structure.
Finally, our findings are limited to one individual and rare structure, which prevent us from generalizing them to the broader population of protoclusters. To identify a larger number of systems similar to MQN01 and study their galaxy populations, it would be necessary to conduct systematic deep and wide surveys, for instance with MUSE or KCWI, targeting extended quasar nebulae as signposts for these structures.


\section{Summary and conclusions}

We detected and studied a population of star-forming galaxies clustered within $\approx4\times4\rm\,cMpc^2$ around the $z\approx3.25$ quasar CTS G18.01, known as MQN01 field. The quasar is surrounded by a giant Ly$\alpha$ emitting nebula discovered by \citet{Borisova2016} and Cantalupo, et al. (in prep.), and the field hosts a large overdensity of dusty star-forming galaxies \citep{Pensabene2024} and AGNs (Travascio, et al. submitted). By leveraging a rich set of multiwavelength data, including MUSE, HST, FORS2, HAWK-I, JWST and {\it Chandra}, we derived the star formation rates and the stellar masses of these galaxies and explored their connection to the local and large-scale overdense environment. The key results are summarized below:

\begin{itemize}
    \item We discovered an overdensity of star forming galaxies of $\rho/\bar{\rho}=53\pm17$ clustered within $4\times4\rm\,cMpc^2$ and $\lvert\Delta\varv\rvert\leq1000\rm\,km\,s^{-1}$ around the central quasar. Measuring the amplitude of the clustering as a function of the projected distance from the quasar, we derived a cross-correlation length that is $\approx3.2$ times larger than the average value found by \citet{Trainor2012} for $z\approx2.7$ quasars. Galaxies in this structure are thus either more clustered or living in a more massive structure than those observed around average quasars at similar times.
    \item We found that the galaxies in the overdensity are forming stars at a rate consistent with the expected main sequence at $z\approx3.25$. In addition, the ability of these galaxies to form stars appears to be more sensitive to the gas and baryon content of the galaxies, probed by their stellar masses, rather than to the local overdensities and the large-scale environment in which they reside.
    \item At the high mass end, the cumulative stellar mass function is flatter compared to the field. This suggests the presence of an excess of massive galaxies in the structure, with $\approx4$ times higher number density than in the field at $\log(M_\star/M_\odot)\geq10.5$. In addition, all the massive galaxies are also X-Ray AGNs. Such an excess of massive star forming galaxies may be due to a higher merger rate or efficient accretion that allow the galaxies to assemble their mass at earlier times or more efficiently than in the field and provide them of enough fuel to continue forming stars at main sequence rates.
    \item The distribution of color-selected Lyman break galaxies on scales $\times36$ larger than the original FoV of the MUSE mosaic suggests that the overdensity extends along the North-South direction and is thus align with the Ly$\alpha$ emitting gas detected by \citet{Borisova2016} around the quasar and to the $\approx4\rm\,cMpc$ filamentary structure observed in emission by Cantalupo et al., (in prep.) This may be a hint that the gaseous structure in which galaxies are embedded extends for tens of cMpc.
\end{itemize}

Overdense structures, such as MQN01, are ideal places to explore the possible connection between the environment and the properties of galaxies at high redshift. The advancement in the study of overdensities at $z>2$ is finally making it possible to explore a wide range of environments, even at the cosmic noon. However, given their rarity, it is still challenging to build a large sample of such structures, and thus to increase the statistical significance of any possible connection between galaxies and environment. As such, even if based on a single field, our study provides important clues for future investigations to piece together the coevolution of galaxies and their large-scale environments from the very early epochs to the late times.

\begin{acknowledgements}
This project was supported by the European Research Council (ERC) Consolidator Grant 864361 (CosmicWeb) and by Fondazione Cariplo grant no. 2020-0902. MiF gratefully acknowledges support from the European Research Council (ERC) under the European Union's Horizon 2020 research and innovation programme (grant agreement no. 757535) and by Fondazione Cariplo, grant no. 2018-2329. This work is based on observations collected at the European Southern Observatory under ESO programmes IDs 102.A-0448(A), and 110.23ZX. This work made use of Astropy (\url{http://www.astropy.org}) a community-developed core Python package and an ecosystem of tools and resources for astronomy \citep{Astropy2013, Astropy2018, Astropy2022}, Numpy \citep{Numpy2020}, SciPy \citep{SciPy2020}, and Matplotlib \citep{Matpltolib2007}.
\end{acknowledgements}

%
%

\bibliographystyle{aa}
\bibliography{Reference_v1}


\begin{appendix}

\section{Custom G-band and UGR selection}
\label{sec:sel_UGR}

\begin{figure} 
\centering
\includegraphics[width=\columnwidth]{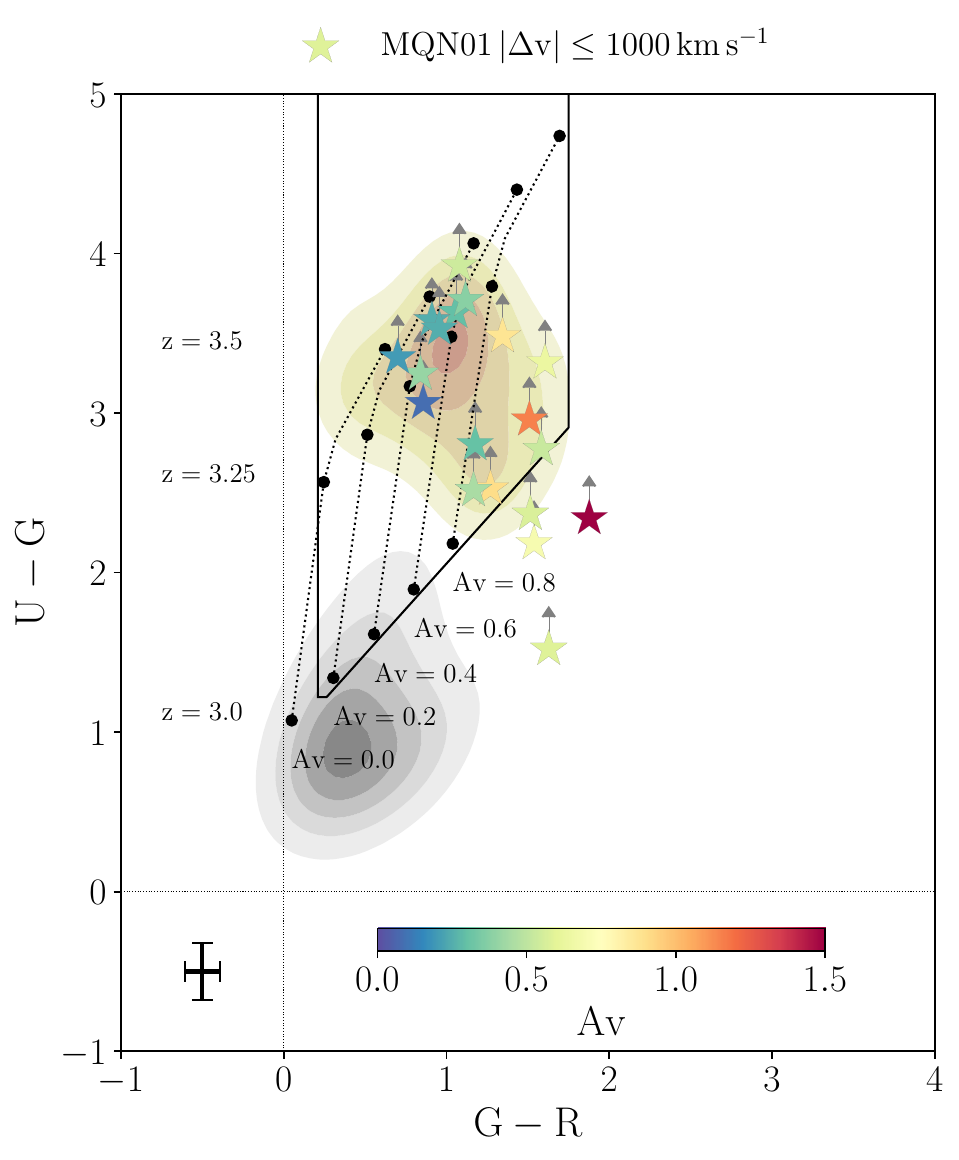}
\caption{Same as Figure \ref{fig:sample_LBG}, but showing the results of the $UGR$ color selection of LBGs. }
\label{fig:sample_LBG_UGR}
\end{figure}

As described in Section \ref{sec:sel_UBR}, we measured $U-B$ and $B-R$ colors to select Lyman break galaxies within the redshift range $z\approx3-3.5$ across a much larger FoV compared to that of the MUSE mosaic. In order to be able to compare our results with other sample in the literature, it is essential to keep the selection function between different studies as similar as possible. Taking advantage of the MUSE data, we restrict to the $4\times4\rm\,cMpc^2$ FoV of the MUSE mosaic and identify the LBG candidates using a $UGR$ (that is, based on $U-G$ and $G-R$ colors) color selection \citep{Steidel1996,Steidel2003}. To this end, we construct a custom-made G-band image covering the FoV of the MUSE mosaic. We therefore combine the B-band image taken with FORS2 with a top-hat medium-band image built out of the MUSE data, producing a custom G-band image with effective wavelength $\lambda_{\rm eff}=4727.3\,$\AA. The MUSE top-hat filter spans a wavelength range from $4920\,$\AA, corresponding to a transmission of $5\%$ for the FORS2 B$_{\rm High}$ filter, to $5510\,$\AA, which corresponds to the maximum wavelength of the FORS2 $\rm G_{\rm High}$ filter.

To calibrate the color selection, we used the same SPS models described in Section \ref{sec:sel_UBR} convolved for the transmission functions of the FORS2 filters and of the custom G-band. As a result, the selection criterion reads as following:

\begin{enumerate}[(i)]
    \item $U>U_{\rm BKG}^{3\sigma}$
    \item $U-G>1.22$
    \item $G-R<0.88\times(U-G)-0.81$
    \item $0.21<G-R<1.75$
\end{enumerate}

Finally, the results of this alternative color selection are shown in Figure \ref{fig:sample_LBG}, while the success rate for the sample of spectroscopically-selected galaxies clustered around the quasar is reported in Table \ref{tab:color_plus_MUSE}. In details, within the $\approx4\times4\rm\,cMpc^2$ FoV of the MUSE mosaic, we are able to select 33 candidate Lyman break galaxies at $z\approx3.0-3.5$, 23 of them also selected by the $UBR$ criterion.

\section{The UV luminosity function of LBGs}
\label{app:LF_LBGs}

\begin{figure} 
\centering
\includegraphics[width=\columnwidth]{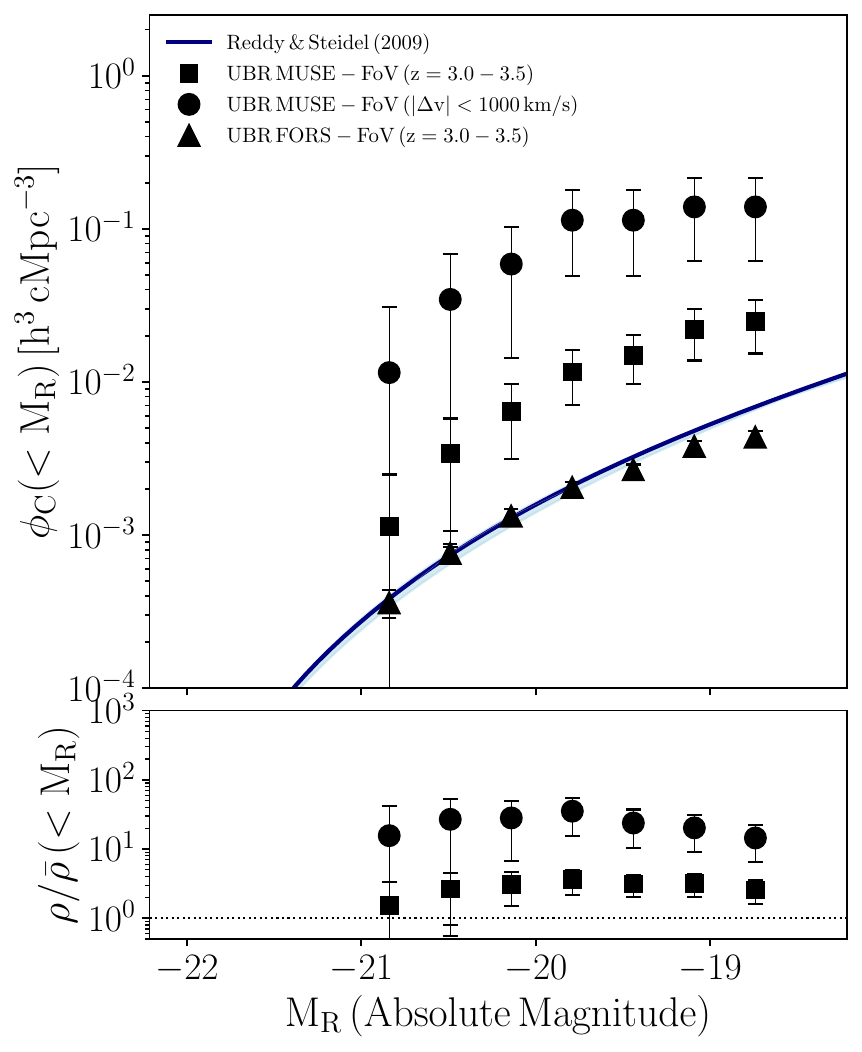}
\caption{Rest-frame UV luminosity function for the sample of $UBR$ color-selected LBGs. Upper panel: cumulative luminositu function of the LBGs identified within the FoV of MUSE (squares) and FORS2 (triangles), and of those galaxies with spectroscopic redshift within $\lvert\Delta\varv\rvert\leq1000\rm\,km\,s^{-1}$ around the central quasar (points). The LBGs' luminosity functions are corrected for the 1D completeness of the R-band that is used for the detection (see Figure \ref{fig:data_FORS}). Lower panel: overdensity as a function of the R-band magnitude.}
\label{fig:LF_LBGs}
\end{figure}

We here estimate the overdensity of $UBR$ color-selected LBGs by applying the same method described in Section \ref{sec:result_LF} and thus deriving the rest-frame UV cumulative luminosity functions in the FoV of MUSE and FORS2 (see Figure \ref{fig:LF_LBGs}). These are corrected for the completeness of the R-band, that is used for the detection of the galaxies (see Figure \ref{fig:data_FORS}). The highest normalization is observed for the LBGs with spectroscopic redshift within $\lvert\Delta\varv\rvert\leq1000\rm\,km\,s^{-1}$ around the quasar in the FoV of the MUSE mosaic. This sample mimics a selection based on photometry and a spectroscopic follow-up, but fails to select massive and dusty galaxies with $\log(M_\star/M_\odot)\gtrsim10$ and $A_{\rm v}\gtrsim0.5$ (as reported in Section \ref{sec:sel_UBR}). In this case, the overdensity estimated relative to the completeness corrected luminosity function from \citet{Reddy2009}, is $\rho/\bar{\rho}=24\pm14$ at $M_{\rm R}\leq-19.25\rm\,mag$ and thus consistent within the uncertainties with the results we obtained for the main spectroscopic sample. An overdensity $\rho/\bar{\rho}=3\pm1$ is observed even when considering a $\times6$ larger volume in redshift. However, the luminosity function becomes fully consistent with \citet{Reddy2009} once computed for all the LBGs detected across the $\times36$ larger area sampled by FORS2, suggesting that the overdensity is diluted in such a large volume. This provides the evidence that the 3D distribution of the star-forming galaxies is significantly peaked not only along the line of sight but also spatially around the quasar. 

\section{Alignment of the LBGs with the large-scale overdensity}
\label{sec:alignement}

\begin{figure} 
\centering
\includegraphics[width=.9\columnwidth]{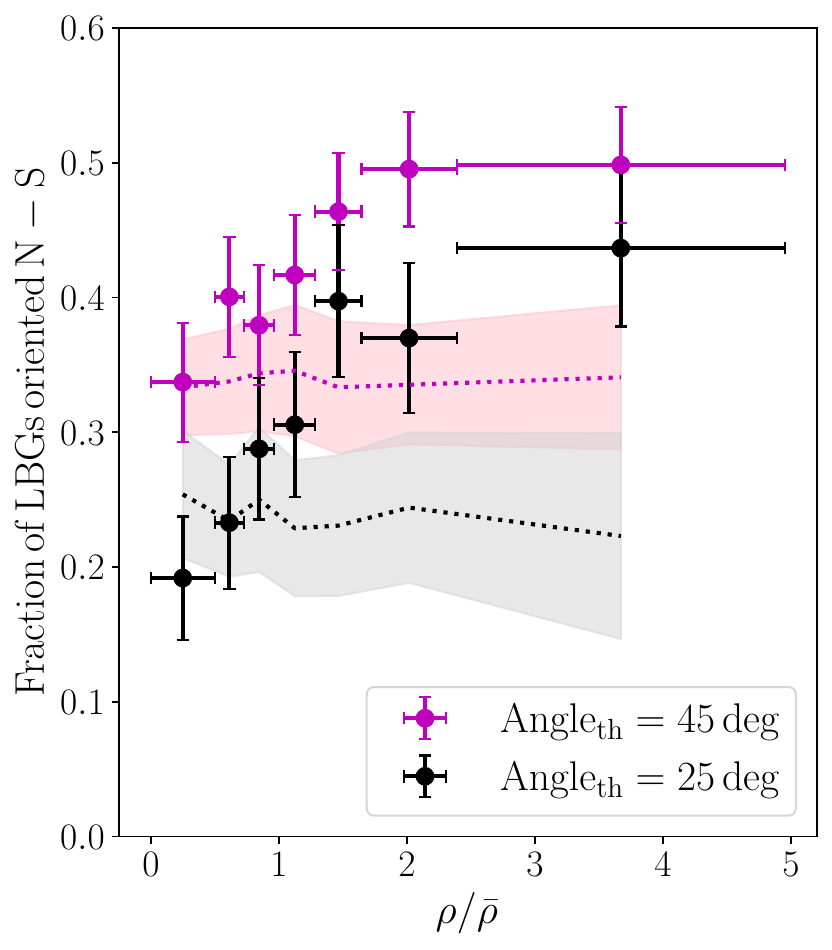}
\caption{Fraction of LBGs aligned with respect to the North-South direction, relative to the quasar, as a function of the local overdensity. We require a galaxy to be aligned if found within $\pm\ang{25}$ (black, corresponding to $\ang{0}-\ang{25}$ or $\ang{155}-\ang{180}$) or $\pm\ang{45}$ (magenta, corresponding to $\ang{0}-\ang{45}$ or $\ang{135}-\ang{180}$) the North-south direction. The shaded area show the 16th and 84th percentiles of the distribution expected for random samples of $1,000$ bootstrap realizations.}
\label{fig:alignement}
\end{figure}

Assuming the LBGs span the entire redshift range $3.0<z<3.5$ (see Section \ref{sec:sel_UBR}), we measured the local co-moving galaxy density by using the $N_{th}$ neighbor estimator with the same formalism presented by \citet{Shimakawa2017}. We choose N=5. Although we expect this method not to be efficient at the edges of the FoV, these galaxies should not affect significantly the statistics and are mostly limited to $R\gtrsim11\rm\,cMpc$. Furthermore, we validate the choice of the estimator by comparing the $N_{th}$ neighbor results to the co-moving density obtained by the scale independent the Voronoi Tessellation, described in Section \ref{sec:over_profiles}. We observe that the 3rd neighbor systematically over-estimate the density by a factor of $\approx1.5$ compared to the Voronoi Tessellation, but the two methods turn out to be consistent for $N\geq5$. The local density is finally converted into a co-moving overdensity $\rho/\bar{\rho}$ against the expected density of galaxies in a blank field (see Section \ref{sec:result_LF}). 

We then measured the orientation of each galaxy with respect to the North-South direction and the central quasar. In bins of co-moving overdensity defined to contain roughly the same number of galaxies, we measured the fraction of LBGs oriented along the North-South direction around the quasar as a function of the environment (see Figure \ref{fig:alignement}). We require a galaxy for being aligned to the North-South direction to be offset by an angle within $\pm\ang{25}$ (black, corresponding to $\ang{0}-\ang{25}$ or $\ang{155}-\ang{180}$) or $\pm\ang{45}$ (magenta, corresponding to $\ang{0}-\ang{45}$ or $\ang{135}-\ang{180}$), chosen so that $\gtrsim40\%$ and $100\%$, respectively, of the MUSE spectroscopic sample is found to be aligned. 

Figure \ref{fig:alignement} shows that the fraction of LBGs oriented along the North-South direction steeply increases with the local co-moving density. In other words, the regions that overdense in LBGs are not sparsely distributed across the large area we surveyed, but instead do preferentially follow the same orientation as the spectroscopically selected galaxies within $4\times4\rm\,cMpc^2$ and $\Delta\varv=\pm1000\rm\,km\,s^{-1}$ from the central quasar. To further support this result, we reproduce the expectations for a sample of randomly-distributed galaxies by injecting the same number of LBGs of the real data in random positions across the FoV. We repeat the injection $1,000$ times and show, despite the large uncertainties derived from the 16th and 84th percentile of the bootstrap distribution, that the fraction of aligned galaxies is flat as a function of the co-moving galaxy overdensity. To support this argument on a more statistical ground, we performed a Spearman's rank correlation test on both the data and the bootstrap results. For the data we found a positive correlation that is significant well above $2\sigma$, as given by $\rho_{\rm spearman}=0.93$, $p-$value$=0.003$ for the threshold $\pm\ang{45}$, and $\rho_{\rm spearman}=0.97$, $p-$value$=0.0005$ for $\pm\ang{25}$. On the other hand, for the random-like samples we find $\rho_{\rm spearman}=0.01$, $p-$value$=0.78$ for the threshold $\pm\ang{45}$, and $\rho_{\rm spearman}=0.02$, $p-$value$=0.86$ for $\pm\ang{25}$. Altogether, this analysis provides a strong evidence for the distribution of massive LBGs to be preferentially aligned along the direction of the overdensity of star-forming galaxies observed in a much smaller volume around the central quasar and thus further supports a picture in which the gaseous structure embedding these galaxies extends for several tens of cMpc.

\section{MQN01 clustering compared to the MAGG survey}
\label{app:clustering_magg}

\begin{figure} 
\centering
\includegraphics[width=\columnwidth]{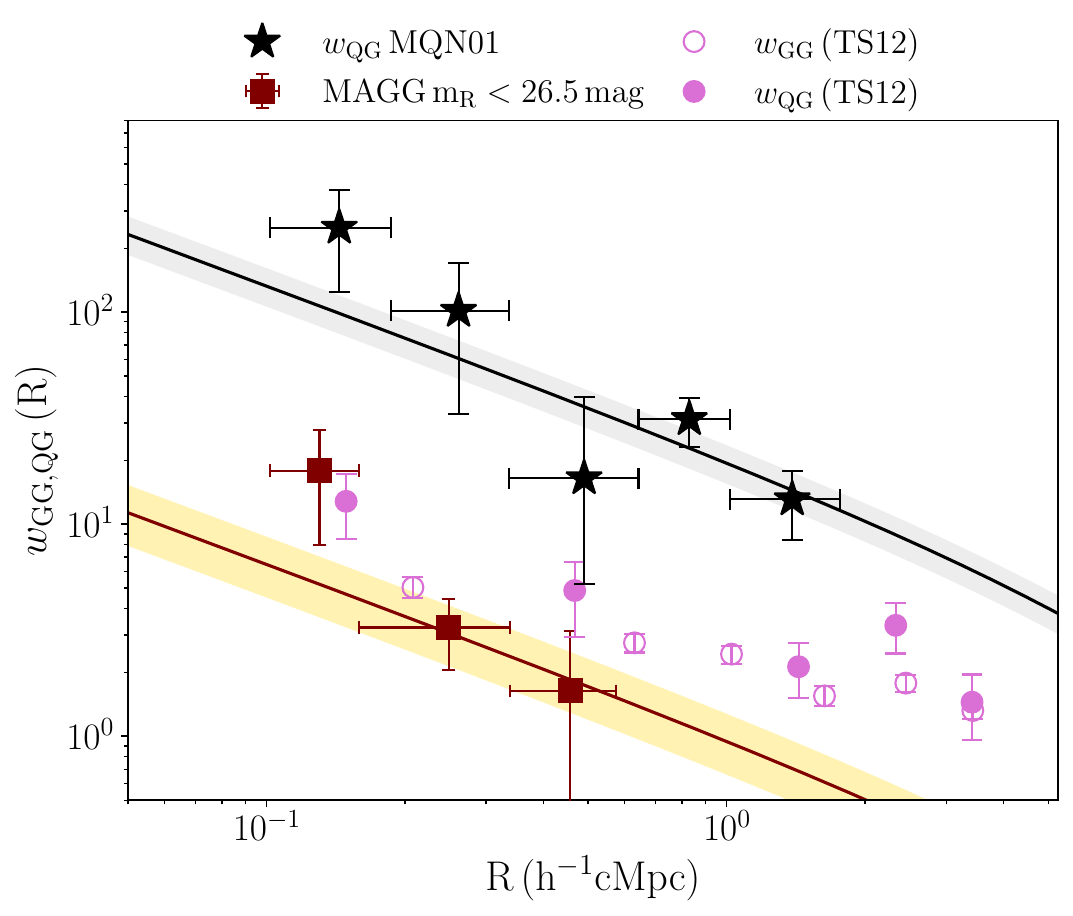}
\caption{Projected quasar-galaxy cross-correlation functions for the continuum-selected galaxies identified in the MAGG survey (darkred and gold-shaded region). Results for the MQN01 galaxies and from TS12 are shown as in Figure \ref{fig:clustering}.}
\label{fig:clustering_magg}
\end{figure}

The clustering analysis presented in Section \ref{sec:radial_overdensity} and discussed in Section \ref{sec:disc_overdensity} provides hints that the MQN01 quasar can possibly be associated to a halo that is much larger compared to those of average quasars at similar redshifts. To further support this conclusion, we here give the details of the quasar-galaxy cross-correlation function we derived for continuum-selected galaxies clustered around the 28 quasars included in the MAGG survey. Such galaxies have been identified by \citet{Lofthouse2020} running {\sc sextractor} \citep{Bertin1996} on the MUSE white-light images obtained for each of the 28 cubes centered on a $3.2\lesssim z\lesssim4.5$ quasar (median $z\approx3.7$). This software outputs a segmentation map that is used to extract the spectra of all the sources, which are finally inspected using {\sc Marz} \citet{Hinton2016} to assign a reliable spectroscopic redshift. As a final step, to produce a sample with a selection function that is as similar as possible to that of the MQN01 galaxies, we selected 20 galaxies with R-band magnitude $m_{\rm R}\lesssim26.5\rm\,mag$, high-quality spectroscopic redshift, and found within a line-of-sight separation $\lvert\Delta\varv\rvert\le1500\rm\,km\,s^{-1}$ from the central quasar of the respective field. We finally derived the projected quasar-galaxy cross-correlation function by using the same procedure described in Section \ref{sec:radial_overdensity}. The results are shown in Figure \ref{fig:clustering_magg}. We measured a cross-correlation length $r_0=(3.8\pm0.8)\,h^{-1}{\rm cMpc}$ for fixed $\gamma=1.5$. Although the different sizes of the FoV, as well as the different selection and average redshift of the samples, make it difficult to compare this result with TS12, the fact that the cross-correlation in MAGG is $\approx6$ times smaller than that of MQN01 remains an important indicator of the rare and massive nature of the quasar host halo, as discussed in Section \ref{sec:disc_overdensity}.

\section{Results from SED fitting}
\label{app:SED_table}

We provide in this section further details and results on the galaxies' SED fitting process. The parameters we set in the initial file of CIGALE are listed below in Table \ref{tab:CIGALE_AGN}, while Table \ref{tab:SED_table} shows the properties estimated for each of the 21 galaxies included in the main spectroscopic sample, as well as their redshift and results of the color-selections employed to identify the LBGs. For the sources detected in X-Ray, the derived SFR and stellar mass may be significantly affected by the contribution of the AGN emission to the total SED. We have thus investigated the morphology of these sources in the JWST/NIRCam data and verified that none of them shows compact bright emission in the center, suggesting that the contribution of the AGN emission does not outshine that of the stellar continuum. Furthermore, according to the best-fit model obtained by running {\sc CIGALE}, the contribution of the AGN emission to the total galaxies' SED is, in terms of flux density, $\gtrsim1\rm\,dex$ smaller compared to the stellar continuum. Based on these considerations, we conclude that the properties we derived for these sources are reliable and are not dominated by presence of an AGN. 

We also tested the results of the SED modeling to be independent of the code we used. To this end, we fit the observed galaxies' photometry using the Monte Carlo Spectro-Photometric Fitter ({\tt MC-SPF}, \citealt{Fossati2018SED,Fossati2019SED}). The grid of stellar spectra is built upon the updated version\footnote{The set of models is publicly available at \url{http://www.bruzual.org/CB19/}. A detailed description is provided in Appendix A of \citet{Sanchez2022}.} of the \citet{BC03} high-resolution models at solar metallicity assuming an exponentially declining star formation history and the \citet{Chabrier2003} initial mass function. A grid of nebular emission lines models from \citet{Byler2018} is scaled to the number of Lyman continuum photons from the stellar models, converted into flux at the redshift of the galaxies and then added to templates. To model the dust attenuation, we assume the double attenuation law from \citet{Calzetti2000}, which includes extra extinction of the young stars compared to those older than 10 Myr. Comparing the two SED fitting codes, the median difference we observe in the galaxies' properties is $0.048\pm0.050\rm\,dex$ for the stellar mass and $0.054\pm0.055\rm\,dex$ for the star formation rate, which is in both the cases substantially smaller than other uncertainties associated to these parameters.

\begin{table*}
\caption{Input parameters for the SED fitting code CIGALE.}
\centering
\begin{tabular}{lcc}
\hline
Parameter & Value \\
\hline
\multicolumn{2}{c}{Exponential or Delayed star formation History} \\
\hline
e-folding time of the main stellar population model (Myr) & 50-15000 in step $\Delta\tau=50\rm\,Myr$ \\
Age of the main stellar population (Myr) & 25-2000 in step $\Delta{\rm age}=5\rm\,Myr$ \\\hline
\multicolumn{2}{c}{Stellar population models: \citet{BC03}} \\
\hline
Initial mass function & \citet{Chabrier2003} \\
Metallicities (in separate runs) & 0.008, 0.004, 0.020 \\
Age of the separation between the young and the old star population (Myr) & 10 \\
\hline
\multicolumn{2}{c}{Dust attenuation law: \citet{Calzetti2000}} \\
\hline
E(B-V)l: color excess of the nebular lines & 0.0-0.75 in step $\rm\Delta E(B-V)l=0.015$ \\
E(B-V)f: reduction factor to compute the E(B-V) \\ for the stellar continuum attenuation & 0.44 \\
\hline
\multicolumn{2}{c}{Nebular Emission: \citet{Inoue2011}} \\
\hline
Ionization parameter & -2.0 \\
Gas metallicity (set to be same as the stellar metallicity) & 0.008, 0.004, 0.020 \\
Fraction of Lyman continuum photons escaping the galaxy & 0.0 \\
\hline
\multicolumn{2}{c}{AGN emission$^{(a)}$: \citet{Fritz2006}} \\
\hline
AGN fraction: contribution of the AGN to the total $L_{\rm IR}$ & 0.0, 0.01, 0.1, 0.2, 0.3, \\ & 0.4, 0.5, 0.6, 0.7, 0.8, 0.9 \\
Opening angle of the dust torus & 10, 30, 60 \\
Angle between AGN axis and line of sight & 0.001, 89.9 \\
Optical depth at 9.7 & 1.0, 6.0 \\
\hline
\multicolumn{2}{c}{X-ray emission$^{(a)}$: \citet{Yang2022_cigale}} \\
\hline
Photon index of the AGN intrinsic X-ray spectrum  & 1.8 \\
Power-law slope connecting $L_{\nu}$ at rest-frame 2500\AA\ and 2keV ($\alpha_{\rm ox}$) & -1.9, -1.7, -1.5, -1.3, -1.1, -0.9 \\
Maximum allowed deviation of $\alpha_{\rm ox}$ from the empirical $\alpha_{\rm ox}-L_{\nu}$ (2500\AA) & 0.2 \\
Deviation from the expected LMXB scaling relation ($\delta_{\rm LMXB}$) & 0.0 \\
Deviation from the expected HMXB scaling relation ($\delta_{\rm HMXB}$) & 0.0 \\
\hline
\end{tabular}
\tablefoot{\tablefoottext{a}{The AGN and X-ray emission are included only for the five sources detected with {\it Chandra}.}}
\label{tab:CIGALE_AGN}
\end{table*}

\begin{table*}
\caption{Detailed results of the galaxies' SED fitting.}
\centering
\begin{tabular}{|c|c|c|c|c|c|c|c|c|}
\hline
Coordinates & spec-z & $\log{[\rm SFR/\rm (M_\odot Yr^{-1})]}$ & $\log{[\rm sSFR/\rm Yr^{-1}]}$ & $\log{[M_\star/M_\odot]}$ & $A_{\rm V}$ & AGN & UBR & UGR \\
\hline
00:41:35.39 -49:37:19.41  & 3.254 & 1.84$\pm$0.30 & -8.62$\pm$0.36 & 10.46$\pm$0.20 & 0.64$\pm$0.03 &   &   & \checkmark  \\
00:41:35.21 -49:37:11.76  & 3.246 & 0.92$\pm$0.30 & -8.61$\pm$0.36 & 9.53 $\pm$0.20 & 0.70$\pm$0.09 &   &   &   \\
00:41:31.95 -49:37:07.89 & 3.253 & 1.02$\pm$0.30 & -8.33$\pm$0.36 & 9.35 $\pm$0.20 & 0.41$\pm$0.03 &   & \checkmark  & \checkmark  \\
00:41:31.60 -49:36:58.25 & 3.245 & 2.26$\pm$0.32 & -8.92$\pm$0.39 & 11.19$\pm$0.22 & 1.15$\pm$0.05 & \checkmark  &   & \checkmark  \\
00:41:26.93 -49:36:50.15 & 3.254 & 1.13$\pm$0.30 & -8.60$\pm$0.36 & 9.74 $\pm$0.20 & 0.30$\pm$0.04 &   & \checkmark  & \checkmark  \\
00:41:34.99 -49:36:41.87 & 3.241 & 1.24$\pm$0.38 & -8.75$\pm$0.44 & 9.99 $\pm$0.23 & 0.90$\pm$0.06 & \checkmark  &   & \checkmark  \\
00:41:32.22 -49:36:39.41 & 3.246 & 1.29$\pm$0.30 & -8.47$\pm$0.36 & 9.76 $\pm$0.20 & 0.53$\pm$0.06 &   &   & \checkmark  \\
00:41:35.04 -49:36:34.90 & 3.236 & 0.32$\pm$0.30 & -8.26$\pm$0.36 & 8.58 $\pm$0.20 & 0.08$\pm$0.02 &   & \checkmark  & \checkmark  \\
00:41:34.50 -49:36:32.13 & 3.243 & 1.11$\pm$0.30 & -8.46$\pm$0.36 & 9.57 $\pm$0.20 & 0.57$\pm$0.06 &   &   &   \\
00:41:32.18 -49:36:20.74 & 3.243 & 1.70$\pm$0.38 & -8.92$\pm$0.43 & 10.62$\pm$0.22 & 1.70$\pm$0.13 & \checkmark  &   &   \\
00:41:31.19 -49:36:20.33 & 3.254 & 0.85$\pm$0.30 & -8.30$\pm$0.36 & 9.15 $\pm$0.20 & 0.25$\pm$0.02 &   & \checkmark  & \checkmark  \\
00:41:31.79 -49:36:18.97 & 3.248 & 0.66$\pm$0.30 & -8.67$\pm$0.36 & 9.33 $\pm$0.20 & 0.46$\pm$0.04 &   &   &   \\
00:41:32.02 -49:36:18.98 & 3.243 & 2.19$\pm$0.33 & -8.89$\pm$0.40 & 11.08$\pm$0.22 & 0.88$\pm$0.06 & \checkmark  &   & \checkmark  \\
00:41:31.92 -49:36:17.84 & 3.250 & 0.27$\pm$0.30 & -9.73$\pm$0.36 & 10.0 $\pm$0.20 & 0.31$\pm$0.25 &   &   &   \\
00:41:34.89 -49:36:15.77 & 3.247 & 0.66$\pm$0.30 & -8.32$\pm$0.36 & 8.97 $\pm$0.20 & 0.30$\pm$0.02 &   & \checkmark  & \checkmark  \\
00:41:34.12 -49:36:02.81 & 3.247 & 1.30$\pm$0.30 & -8.26$\pm$0.36 & 9.56 $\pm$0.20 & 0.38$\pm$0.02 &   & \checkmark  & \checkmark  \\
00:41:30.93 -49:35:39.02 & 3.239 & 0.96$\pm$0.30 & -8.24$\pm$0.36 & 9.20 $\pm$0.20 & 0.25$\pm$0.02 &   & \checkmark  & \checkmark  \\
00:41:30.94 -49:35:34.61 & 3.248 & 1.61$\pm$0.30 & -8.30$\pm$0.36 & 9.91 $\pm$0.20 & 0.54$\pm$0.03 &   & \checkmark  & \checkmark  \\
00:41:29.98 -49:35:34.32 & 3.258 & 0.71$\pm$0.30 & -8.22$\pm$0.36 & 8.93 $\pm$0.20 & 0.20$\pm$0.03 &   & \checkmark  & \checkmark  \\
00:41:32.25 -49:35:29.16 & 3.244 & 0.01$\pm$0.30 & -8.96$\pm$0.36 & 9.00 $\pm$0.20 & 0.45$\pm$0.15 &   & \checkmark  & \checkmark  \\
00:41:35.13 -49:37:12.81 & 3.245 & 2.09$\pm$0.30 & -9.28$\pm$0.36 & 11.37$\pm$0.20 & 0.58$\pm$0.03 & \checkmark  &   &   \\
\hline
\end{tabular}
\tablefoot{The properties listed in the table refer to the sample of spectroscopically selected galaxies that have been identified in the FoV of the MUSE mosaic within a line-of-sight separation $\lvert\Delta\varv\rvert\leq1000\rm\,km\,s^{-1}$ from the central quasar. Uncertainties on the SFR, sSFR and $M_\star$ are computed as described in Section \ref{sec:SED_fitting}.}
\label{tab:SED_table}
\end{table*}

\end{appendix}

\end{document}